\documentclass[11pt,twoside]{article}

\usepackage{amsmath,amsfonts,amssymb,amsthm}
\usepackage{epsfig}
\newcommand{\rd}{{\rm d}}

\newcommand{\field}[1]{\mathbb{#1}}
\newcommand{\N}{\field{N}}
\newcommand{\R}{\field{R}}
\newcommand{\C}{\field{C}}
\newcommand{\D}{\mathcal{D}}
\newcommand{\E}{\mathcal{E}}
\newcommand{\F}{\mathcal{F}}
\renewcommand{\H}{\mathcal{H}}
\newcommand{\h}{\mathfrak{h}}
\renewcommand{\L}{\mathbf{B}}
\newcommand{\HxF}{\tilde{\mathcal{H}}}   

\newcommand{\Hgs}{\mathcal{H}_{\text{das}}} 
\newcommand{\Pgs}{P_{\text{das}}}           

\newcommand{\Hmod}{H_{\text{mod}}}         
\newcommand{\Hmodex}{\tilde{H}_{\text{mod}}}

\newcommand{\Hex}{\tilde{H}}
\newcommand{\uGamma}{\breve{\Gamma}}

\newcommand{\dGamma}{\mathrm{d}\Gamma}

\newcommand{\eps}{\varepsilon}
\newcommand{\ph}{\varphi}

\newcommand{\ran}{\mathrm{Ran}}

\newcommand{\restricted}{|\grave{}\,}

\newcommand{\Hato}{H_{\text{at}}}
\newcommand{\Hatom}{H_{\text{atom}}}

\newcommand{\sprod}[2]{\mbox{$\langle #1,#2 \rangle$}}     

\newcommand{\Ran}{\operatorname{Ran}}
\newcommand{\supp}{\operatorname{supp}}

\newtheorem{theorem}{Theorem}
\newtheorem{lemma}[theorem]{Lemma}
\newtheorem{corollary}[theorem]{Corollary}

\newtheorem{prop}[theorem]{Proposition}

\begin{document}
\title{\bf Rayleigh Scattering at Atoms with Dynamical Nuclei}
\author{\vspace{5pt} J. Fr\"ohlich$^1$\footnote{E-mail:
juerg@itp.phys.ethz.ch. Activities supported, in part, by a grant
from the Swiss National Foundation.} , M. Griesemer$^2$\footnote{
E-mail: marcel@mathematik.uni-stuttgart.de. Work partially
supported by U.S. National Science Foundation grant DMS 01-00160.
} and B. Schlein$^3$\footnote{E-mail: schlein@math.stanford.edu. 
Supported by a NSF postdoctoral fellowship.} \\
\vspace{-4pt}\small{$1.$ Theoretical Physics, ETH--H\"onggerberg,} \\
\small{CH--8093 Z\"urich, Switzerland}\\
\vspace{-4pt} \small{$2.$ Fachbereich Mathematik,
Universit\"at Stuttgart,} \\ \small{D-70569 Stuttgart, Deutschland}\\
\vspace{-4pt}
\small{$3.$ Department of Mathematics, Harvard University,} \\
\small{Cambridge, MA 02138, USA}\\}

\maketitle

\begin{abstract}
Scattering of photons at an atom with a {\it dynamical} nucleus is studied on
the subspace of states of the system with a total energy below the threshold
for ionization of the atom ({\it Rayleigh scattering}). The kinematics of the
electron and the nucleus is chosen to be non-relativistic, and their spins are
neglected. In a simplified model of a hydrogen atom or a one-electron ion
interacting with the quantized radiation field in which the helicity of
photons is neglected and the interactions between photons and the electron and
nucleus are turned off at very high photon energies and at photon energies
below an {\it arbitrarily small}, but {\it fixed} energy (infrared cutoff),
{\it asymptotic completeness of Rayleigh scattering} is established
rigorously. On the way towards proving this result, it is shown that, after
coupling the electron and the nucleus to the photons, the atom still has a
stable ground state, provided its center of mass velocity is smaller than the
velocity of light; but its excited states are turned into resonances. The
proof of asymptotic completeness then follows from extensions of a positive
commutator method and of propagation estimates for the atom and the photons
developed in previous papers.

The methods developed in this paper can be extended to more realistic
models. It is, however, not known, at present, how to remove the infrared
cutoff.
\end{abstract}

\section{Introduction}\label{sec:intro}

During the past decade, there have been important advances in our
understanding of the mathematical foundations of quantum electrodynamics with
non-relativistic, quantum-mechanical matter (``non-relativistic QED''). Subtle
spectral properties of the Hamiltonians generating the time evolution of 
atoms and molecules interacting with the quantized radiation field have been
established. In particular, existence of atomic ground states and 
absence of stable excited states have been proven, and the energies and life
times of resonances have been calculated in a rigorously controlled way, for a
variety of models; see \cite{BFS1,BFSS,HS,Sk,GLL,LL,G,AGG,BFP}, and references
given there. Furthermore, some important steps towards developing the
scattering theory for systems of non-relativistic matter interacting with
massless bosons, in particular photons, have been taken. Asymptotic
electromagnetic field operators have been constructed in \cite{FGS1}, and wave
operators for Compton scattering have been shown to exist in
\cite{P}. Rayleigh scattering, i.e., the scattering of photons at atoms below
their ionization threshold, has been analyzed in \cite{FGS2} for models with an
infrared cutoff. The results in this paper are based on methods developed in
\cite{DG}. Some earlier results on Rayleigh scattering have been derived in
\cite{S}. Compton scattering in models with an infrared cutoff has been
studied in \cite{FGS3}. While it is understood how to calculate scattering
amplitudes for various low-energy scattering processes in models {\it without}
an infrared cutoff, all known {\it general} methods to prove {\it unitarity of
  the scattering matrix} on subspaces of states of sufficiently low energy,
i.e., {\it asymptotic completeness}, require the presence of an (arbitrarily
small, but {\it positive}) infrared cutoff. In \cite{S,DG,FGS2}, Rayleigh
scattering has only been studied for models of atoms with {\it static} (i.e.,
{\it infinitely heavy}) nuclei.

As suggested by this discussion, the main challenge in the scattering theory
for non-relativistic QED presently consists in solving the following problems:
\begin{enumerate}
\item[i)] To remove the infrared cutoff in the analysis of Rayleigh scattering;
\item[ii)] to remove the infrared cutoff in the treatment of Compton
  scattering of photons at one freely moving electron or ion;
\item[iii)] to prove {\it asymptotic completeness} of Rayleigh scattering of
  photons at atoms or molecules with {\it dynamical} nuclei.
\end{enumerate}

In this paper, we solve problem iii) in the presence of an (arbitrarily small,
but positive) infrared cutoff. In order to render our analysis, which is quite
technical, as simple and transparent as possible, we consider the simplest
model exhibiting all typical features and challenges encountered in an
analysis of problem iii). We consider a hydrogen atom or a one-electron
ion. The nucleus is described, like the electron, as a non-relativistic
quantum-mechanical point particle of {\it finite} mass. We ignore the spin
degrees of freedom of the electron and the nucleus. The electron and the
nucleus interact with each other through an attractive two-body potential $V$,
which can be chosen to be the electrostatic Coulomb potential. Electron and
nucleus are coupled to a quantized radiation field. As in \cite{FGS2}, the
field quanta of the radiation field are massless bosons, which we will call
photons. Physically, the radiation field is the quantized electromagnetic
field. However, the helicity of the photons does not play an interesting role
in our analysis, and we therefore consider {\it scalar} bosons.

As announced, we focus our attention on Rayleigh scattering; i.e., we only
consider the asymptotic dynamics of states of the atom and the radiation field
with energies below the threshold for break-up of the atom into a freely
moving nucleus and electron, i.e., below the ionization threshold. Moreover,
we introduce an {\it infrared cutoff}: Photons with an energy below a certain
arbitrarily small, but positive threshold energy do {\it not} interact with
the nucleus and the electron. While all our other simplifications are purely
cosmetic, the presence of an infrared cutoff is crucial in our proof of
asymptotic completeness (but {\it not} for most other results presented in
this paper).

In the model we study, the atom can be located anywhere in physical space and
can move around freely, and the Hamiltonian of the system is
translation-invariant. This feature suggests to combine and extend the
techniques in two previous papers, \cite{FGS2} (Rayleigh scattering with
static nuclei) and \cite{FGS3} (Compton scattering of photons at freely moving
electrons), and this is, in fact, the strategy followed in the present
paper. As in \cite{FGS3}, to prove asymptotic completeness 
we must impose an upper bound on the energy of the
state of the system that guarantees that the center of mass velocity of the
atom does not exceed one third of the velocity of light. This is a purely
technical restriction which we believe can be replaced by one that guarantees
that the velocity of the atom is less than the velocity of light. We note
that, for realistic atoms, the condition that the total energy of a state be
below the ionization threshold of an atom at rest automatically guarantees
that the speed of the atom in an arbitrary internal state is much less than a
third of the speed of light.

Next, we describe the model studied in this paper more explicitly. The Hilbert
space of pure states is given by
\begin{equation}
\H = L^2 (\R^3, \rd x_n) \otimes L^2 (\R^3, \rd x_e) \otimes \F,
\end{equation}
where the variables $x_n$ and $x_e$ are the positions of the nucleus and of
the electron, respectively, and $\F$ is the symmetric Fock space over the
one-photon Hilbert space $L^2 (\R^3, \rd k)$, where the variable $k$ denotes
the momentum of a photon. Vectors in $\F$ describe pure states of the
radiation field.

The Hamiltonian generating the time evolution of states of the system is given
by
\begin{equation}\label{eq:ham1}
H_g = \Hatom + H_f + g \left( \phi(G^e_{x_e}) + \phi (G^n_{x_n})\right)
\end{equation}
where
\begin{equation}
\Hatom = \frac{p_e^2}{2m_e} + \frac{p_n^2}{2m_n} + V(x_e -x_n)
\end{equation}
is the Hamiltonian of the atom decoupled from the radiation field. Here
$p_e=-i\nabla_{x_e}$ and $p_n = -i\nabla_{x_n}$ are the momentum
operators of the electron and the nucleus, respectively, and $m_e$
and $m_n$ are their masses. The term $V(x_e-x_n)$ is the potential of an
attractive two-body force, e.g., the electrostatic Coulomb force, between the
electron and the nucleus; ($V(x)$ is negative and tends to zero, as $|x| \to
\infty$, and it is assumed to be such that the spectrum of $\Hatom$ is bounded
from below and has at least one negative eigenvalue). The operator $H_f$ on
the r.h.s. of (\ref{eq:ham1}) is the Hamiltonian of the free radiation
field. It is given by
\[ H_f = \int \rd k \,
|k| a^* (k) a(k),\] where $|k|$ is the energy of a photon with momentum $k$,
and $a^* (k)$, $a(k)$ are the usual
boson creation- and annihilation operators: For every function
$f \in L^2 (\R^3, \rd k)$,
\[ a^* (f) = \int \rd k \, f(k) a^* (k) \qquad  \text{and} \qquad
a(f) = \int \rd
k\, \overline{f(k)} a(k)\, \] are densely defined, closed operators on the
Fock space $\F$, and, for $f,h, \in L^2 (\R^3, \rd k)$, they satisfy the
canonical commutation relations
\[ [a(f) , a^* (h) ] = (f,h), \quad [a^{\sharp} (f) ,
a^{\sharp} (h)] = 0 ,
\] where $(f,h)$ denotes the scalar product of $f$ and $h$.
For $f\in L^2(\R^3,\rd k)$, a field operator $\phi(f)$ is defined
by
\begin{equation*}\label{eq5}
   \phi(f) = a(f) + a^*(f).
\end{equation*}
It is a densely defined, self-adjoint operator on $\F$. The
functions (form factors) $G_{x_e}^e$ and $G_{x_n}^n$ on the right
side of \eqref{eq:ham1} are given by
\begin{equation}\label{eq6}
   G_{x_e}^e (k) = e^{-ik\cdot x_e}\kappa_e(k),\qquad G_{x_n}^n (k)
=e^{-ik\cdot x_n}\kappa_n(k),
\end{equation}
where $\kappa_e$ and $\kappa_n$ belong to the Schwartz space, and
\begin{equation*}\label{eq7}
   \kappa_e(k)=\kappa_n(k)=0, \qquad\ \text{for all}\ k\in \R^3 
\text{ with}\ |k|\leq \sigma,
\end{equation*}
for some $\sigma>0$ (infrared cutoff). In many of our results, we could pass to
the limit $\sigma=0$; but in our proof of asymptotic completeness of Rayleigh
scattering, the condition that $\sigma>0$ is essential. Finally, the parameter
$g$ on r.h.s. of (\ref{eq:ham1})
is a coupling constant; it is assumed to be non-negative and will
be chosen sufficiently small in the proofs of our results. (It should be noted
that we are using units such that Planck's constant $\hbar=1$ and the speed of
light $c=1$, and we work with dimensionless variables $x_e,x_n,k$ chosen such
that $\Hatom$ and $H_f$ are independent of $g$.)

In the description of the atom, it is natural to use the following variables:
\[
   X=\frac{m_e x_e+m_n x_n}{m_e+m_n}, \qquad x=x_e-x_n.
\]
Here $X$ is the position of the center of mass of the atom, and $x$ is the
position of the electron relative to the one of the nucleus. Then
\begin{equation}\label{eq8}
   H_g =  \frac{P^2}{2M} + \frac{p^2}{2m} + V(x) + H_f +
   g\left(\phi(G^e_{X+\frac{m_n}{M} x}) + \phi(G^n_{X-\frac{m_e}{M}x})\right),
\end{equation}
where $P=-i\nabla_X$, $p=-i\nabla_x$, $M=m_e+m_n$, and $m=m_e m_n M^{-1}$;
(center-of-mass momentum, relative momentum, total mass, reduced mass,
respectively). Self-adjointness of $H_g$ on $\H$ (under appropriate assumptions
on $V$) is a standard result.

In this paper, we study the dynamics generated by $H_g$ on the subspace of
states in $\H$ whose maximal energy is below the ionization threshold
 \begin{equation}\label{eq9}
   \Sigma_{\text{ion}} = \lim_{R \to \infty} \inf_{\ph \in \D_R}
\frac{\sprod{\ph}{H_g \ph}}{\sprod{\ph}{\ph}},
\end{equation}
where \( \D_R = \{\ph \in \H : \chi (|x|\geq R)\ph=\ph \} \) is
the subspace of vectors  in $\H$ with the property that the
distance between the electron and the nucleus is at least $R$.
Vectors in $\H$ with a maximal total energy {\it below}
$\Sigma_{\text{ion}}$ exhibit exponential decay in $|x|$, the
distance between the electron and the nucleus; see \cite{G}. Under our
assumptions, $-Cg^2\leq \Sigma_{\text{ion}} \leq 0$, for a finite
constant $C$ depending only on $\kappa_e$ and $\kappa_n$. When $g
\downarrow 0$ then $\Sigma_{\text{ion}} \uparrow 0$, which is the
ionization threshold of a one-electron atom or -ion decoupled from
the radiation field.

Our choice of the Hamiltonian $H_g$, see \eqref{eq:ham1} and \eqref{eq8},
and of the form factors $G^e_{x_e}$ and $G^n_{x_n}$, see \eqref{eq6},
makes it clear
that the dynamics of the system is space-translation invariant: Let
\begin{equation}\label{eq10}
   P_f = \int\rd k \, k  a^*(k) a(k)
\end{equation}
denote the momentum operator of the radiation field, and let $\Pi=P+P_f$
be the total momentum operator. It is easy to check that
\begin{equation}\label{eq11}
    [H_g,\Pi] = 0\qquad \text{(translation invariance)}.
\end{equation}
It is then useful to consider direct-integral decompositions of the space
$\H$ and the operator $H_g$ over the spectrum of the total momentum operator
$\Pi$ (which, as a set, is $\R^3$). Thus
\begin{equation}\label{eq12}
   \H=\int_{\R^3}^{\oplus}\rd \Pi \, \H_{\Pi}, \quad \text{with} \quad
\H_{\Pi}\simeq  L^2(\R^3,dx) \otimes \F,
\end{equation}
and
\begin{equation}
  H_g=\int_{\R^3}^{\oplus}\rd \Pi \, H_g (\Pi),
\end{equation}
where the fiber Hamiltonian $H_g (\Pi)$ is the operator on the fiber Hilbert
space $\H_{\Pi}$ given by
\begin{equation}\label{eq13}
   H_g(\Pi) = \frac{(\Pi-P_f)^2}{2M}+ \Hato + H_f + 
g\left(\phi(G^e_{\frac{m_n}{M}x})
  +\phi(G^n_{-\frac{m_e}{M}x})\right),
\end{equation}
where $\Pi-P_f$ is the center-of-mass momentum of the atom, and
$\Hato=p^2 / 2m +V(x)$ is the Hamiltonian describing the relative
motion of the electron around the nucleus.

We are now in the position to summarize the main results proven in this paper
for the model introduced above. In a first part, we analyze the energy
spectra of the fiber Hamiltonians $H_g(\Pi)$ below a certain threshold
$\Sigma< \min (\Sigma_{\text{ion}}, \Sigma_{\beta})$,
where $\Sigma_{\text{ion}}$ is given in
\eqref{eq9}, and $\Sigma_{\beta} =E_0^{\text{at}} + M\beta^2/2$;
$E_0^{\text{at}}$ is the ground state energy of
$\Hato$, and $\beta$ is a constant $<1$
($=$speed of light, in our units). The condition $\Sigma<\Sigma_{\text{ion}}$
guarantees that the electron is bound to the nucleus, with exponential decay
in $x$, and $\Sigma<\Sigma_{\beta<1}$ implies that, for a sufficiently small
coupling constant $g$, the center-of-mass velocity of the atom is smaller than
the speed of light. (For center-of-mass velocities $>1$, the ground state
energy of the atom decoupled from the radiation field is embedded in
continuous spectrum, and the ground state becomes unstable when the coupling
to the radiation field is turned on.) For realistic atoms, in particular for
hydrogen, $\Sigma_{\text{ion}} \ll \Sigma_{\beta=1/4}$, so that
$\Sigma<\Sigma_{\text{ion}}$ is the only relevant condition.
We let $E_g(\Pi)=\inf\sigma(H_g(\Pi))$ denote
the ground state energy of $H_g(\Pi)$, and we define
\begin{equation}\label{eq14}
  B_{\Sigma} = \{\Pi\in \R^3 : \; E_0^{\text{at}}
+\frac{\Pi^2}{2M} \leq \Sigma\}.
\end{equation}

We prove that, for every $\Pi\in B_{\Sigma}$, $E_g(\Pi)$ is a
simple eigenvalue of $H_g(\Pi)$, i.e., that the atom has a
\emph{unique ground state}, provided $g$ is small enough. This is
a result that is expected to survive the limit $\sigma \downarrow
0$, provided the factors $\kappa_e$ and $\kappa_n$ are not too
singular at $k=0$. For the Pauli-Fierz model of non-relativistic
QED, existence of a ground state can be proven under conditions
similar to the ones described above, {\it provided} the total charge of
electrons and nucleus vanishes; see \cite{AGG}.

By appropriately modifying Mourre theory in a form developed in
\cite{BFSS}, we prove that the spectrum of $H_g(\Pi)$ in the
interval $(E_g(\Pi),\Sigma)$ is purely continuous. With relatively
little further effort, our methods would also show that $\sigma
(H_g(\Pi))\cap (E_g(\Pi),\Sigma)$ is absolutely continuous. (These
results, too, would survive the removal of the infrared cutoff,
$\sigma\downarrow 0$. This will not be shown in this paper; but
see \cite{FGSi}.)

We denote the ground state of $H_g(\Pi),\ \Pi\in B_{\Sigma}$, by $\psi_{\Pi}$;
($\psi_{\Pi}$ is called the \emph{dressed atom} (ground) \emph{state} of
momentum $\Pi$). The space of wave packets of dressed atom states, $\Hgs$,
is the subspace of the total Hilbert space $\H$ given by
\begin{equation}\label{eq15}
  \Hgs = \left\{\psi(f): \; \psi(f)=\int^{\oplus}\rd \Pi
\, f(\Pi) \psi_{\Pi},\ f\in L^2(B_{\Sigma}, \rd \Pi) \right\}.
\end{equation}
This space is invariant under the time evolution. In fact, 
$e^{-iH_g t}\psi(f)=\psi(f_t)$, where, for $f\in L^2
(B_{\Sigma},\rd\Pi)$,
\(f_t(\Pi)= e^{-iE_g(\Pi)t}f(\Pi)\in L^2(B_{\Sigma},\rd\Pi) \), for all
times $t$.

In a second part of our paper, scattering theory is developed for the models
introduced above. We first construct asymptotic photon creation- and
annihilation operators
\begin{equation}\label{eq16}
  a_{\pm}^{\sharp}(h)\ph = s-\lim_{t\to \pm\infty} e^{iH_g t}a^{\sharp}(h_t)
  e^{-iH_g t}\ph,
\end{equation}
where $h_t(k) = e^{-i|k|t}h(k)$ is the free time evolution of a
one-photon state $h(k)$. To ensure the existence of the strong
limit on the r.h.s. of \eqref{eq16}, we assume that $h\in
L^2(\R^3,(1+|k|^{-1})\rd k)$, that $\ph$ belongs to the range of
the spectral projection, $E_{\Sigma}(H_g)$, of $H_g$ corresponding
to the interval $(-\infty,\Sigma]$, with
$\Sigma<(\Sigma_{\text{ion}},\Sigma_{\beta})$, as above, and
$\beta<1$, and that the coupling constant $g$ is so small
(depending on $\Sigma$) that the velocity of the center of mass of
the atom is smaller than one. The last condition ensures that the
distance between the atom and a configuration of outgoing photons
increases to infinity and hence the interaction between these
photons and the atom tends to $0$, as time $t$ tends to $+\infty$.
The details of the proof of \eqref{eq16} are very similar to those
in \cite{FGS1}. From \eqref{eq15} and \eqref{eq16} we infer that,
for $\psi(f)\in \Hgs$, and under the conditions of existence of
the limit in \eqref{eq16}, vectors of the form
\(a_{\pm}^{*}(h_1)\cdots a_{\pm}^{*}(h_n)\psi(f)\) exist, and
their time evolution is the one of freely moving photons and a
freely moving atom:
\begin{equation}\label{eq17}
e^{-iH_gt} a_{\pm}^{*}(h_1)\dots a_{\pm}^{*}(h_n)\psi(f) =
a^{*}(h_{1,t})\cdots a^{*}(h_{n,t})\psi(f_t) + o(1),
\end{equation}
as $t\to\pm\infty$. Furthermore, $a_{\pm}(h)\psi(f)=0$, under the same
assumptions. Equation \eqref{eq17} provides the justification for calling the
vectors \(a_{\pm}^{*}(h_1)\dots a_{\pm}^{*}(h_n)\psi(f)\) {\it scattering
states}. We already know that the atom does {\it not} have any stable excited
states. It is therefore natural to expect that the time evolution of an
arbitrary vector in the range of the spectral projection $E_{\Sigma}
(H_g)$, with $\Sigma<\min(\Sigma_{\text{ion}}, \Sigma_{\beta<1})$
as above, approaches a vector
describing a configuration of freely moving photons and a freely moving atom
in its ground state, as time $t$ tends to $\pm \infty$. Thus, with
\eqref{eq17} and \eqref{eq15}, we expect that, for
$\Sigma<\min(\Sigma_{\text{ion}}, \Sigma_{\beta<1})$, 
\begin{multline}\label{eq18}
\Big\langle \left\{ a_{\pm}^* (h_1) \dots a_{\pm}^* (h_n) \psi (f)
: \right. \\ \left. \psi (f) \in \Hgs,
h_j \in L^2 (\R^3, (1+1/|k|) \rd k), j=1,\dots,n, n=1,2,\dots
\right\} \Big\rangle^- \\ \supset \ran
E_{\Sigma}(H_g) \,,
\end{multline}
where $\langle S \rangle$ denotes the linear subspace spanned by a set, $S$, of
vectors in $\H$, and $\langle S \rangle^{-}$ denotes the closure of $\langle S
\rangle$ in the norm of $\H$. Property \eqref{eq18} is called {\it asymptotic
completeness of Rayleigh scattering}. The {\it main result} of this paper is a
proof of \eqref{eq18} under the supplementary condition 
that $\Sigma < \Sigma_{\beta}$ for some $\beta <1/3$  
(the proof of \eqref{eq18} is the only part of the paper where, 
for technical reasons, we need to assume $\beta
<1/3$; all other results only require $\beta <1$). 
Next, we reformulate \eqref{eq18} in a more convenient language. We define a
Hilbert space of scattering states as the space $\Hgs \otimes \F$, and we
introduce an asymptotic Hamilton operator, $\tilde{H}^{\rm das}_{g}$, by 
setting
\begin{equation}\label{eq19}
   \tilde{H}^{\rm das}_{g} = H^{\rm das}_{g}\otimes 1 + 1\otimes H_f ,
\end{equation}
where
\begin{equation}
  H^{\rm das}_{g}\psi(f) = \psi(E_g( . \, )f),
\end{equation}
for arbitrary $f\in L^2(B_{\Sigma},\rd \Pi)$, with $\Sigma<\min
( \Sigma_{\text{ion}}, \Sigma_{\beta<1/3})$, as above.
On the range of $E_{\Sigma} (\tilde{H}^{\rm das}_{g})$, the
operators $\Omega_{\pm}$, given by
\begin{equation}\label{eq20}
  \Omega_{\pm}(\psi(f)\otimes a^{*}(h_1)\dots a^{*}(h_n)\Omega)
= a_{\pm}^{*}(h_{1})\dots a_{\pm}^{*}(h_{n})\psi(f)
\end{equation}
exist; see \eqref{eq16}. The vector $\Omega$ is the vacuum in the Fock-space
characterized by the property that $a(h)\Omega = 0$, for $h\in L^2(\R^3,\rd
k)$. The operators $\Omega_{+}$ and $\Omega_{-}$ are called {\it
wave operators}, and the {\it scattering matrix} is defined by
\begin{equation}\label{eq21}
    S = \Omega_{+}^{*}\Omega_{-}.
\end{equation}
{F}rom Eqs. \eqref{eq16} and \eqref{eq17} we find that
\begin{equation*}
    e^{-iH_g t}\Omega_{\pm} = \Omega_{\pm} e^{-i\tilde{H}^{\rm das}_{g}t},
\end{equation*}
and hence the ranges of $\Omega_{+}$ and  $\Omega_{-}$ are contained in the
range of $E_{\Sigma}(H_g)$. Using that $a_{\pm} (h) \psi (f) =0$, for $h \in
L^2 (\R^3,\rd k)$ and $\psi (f) \in \Hgs$, one sees that $\Omega_+$ and
$\Omega_-$ are isometries from the range of $E_{\Sigma} (\tilde{H}^{\rm das}_{g})$ into $\H$.
If we succeeded in proving that
\begin{equation}\label{eq22}
    \ran \left(\Omega_{\pm} \restricted{\ran E_{\Sigma}(\tilde{H}^{\rm das}_{g})}\right) =
\ran E_{\Sigma}(H_g)
\end{equation}
we would have established the unitarity of the $S$-matrix, defined in
\eqref{eq21}, on $\ran E_{\Sigma}(\tilde{H}^{\rm das})$, i.e., asymptotic completeness of
Rayleigh scattering.

In order to prove (\ref{eq22}), we show that $\Omega_{\pm}$ have right
inverses defined on $\ran E_{\Sigma} (H_g)$. Our proof is inspired by proofs
of similar results in \cite{DG} and in \cite{FGS2,FGS3}. It is based on
constructing a so called {\it asymptotic observable} $W$ and then proving that
$W$ is {\it positive} on the orthogonal complement of $\Hgs$ in $E_{\Sigma}
(H_g)$. The proof of this last result is, perhaps, the most original
accomplishment in this paper and is based on some new ideas.

Our paper is organized as follows. In Section \ref{sec:model}, we
define our model more precisely, and we state our assumptions on the
potential $V(x)$ and on the form factors $G^e_{x_e}$ and
$G^n_{x_n}$. In Section \ref{sec:spectrum}, we study the spectrum
of the fiber Hamiltonian $H_g(\Pi)$: In Section \ref{sec:das}, we
prove the existence of dressed atom states, and, in Section
\ref{sec:poscomm}, we prove two positive commutator estimates, from
which we conclude that the spectrum of $H_g(\Pi)$ above the ground state
energy and below an appropriate threshold is continuous. In
Section \ref{sec:scatt}, we discuss the scattering theory of the
system. First, in Section \ref{sec:waveop}, we prove the existence
of asymptotic field operators, we recall some of their
properties, we prove the existence of the wave operators, and we
state our main theorem. Then, in Section \ref{sec:modham}, we
introduce a modified Hamiltonian, $\Hmod$, describing ``massive'' photons,
and we explain why it
is enough to prove asymptotic completeness for $\Hmod$ instead of
$H_g$. In Section \ref{sec:observable}, we construct asymptotic
observables $W$ and inverse wave operators $W_{\pm}$. In
Section \ref{sec:pos}, we prove positivity of our asymptotic
observables when restricted to the orthogonal complement of
$\H_{\text{das}}$ (the space of wave packets of dressed atom
states). Finally, in Section \ref{sec:AC}, we complete the proof of
asymptotic completeness. In Appendix \ref{sec:Fock}, we introduce some
notation, used throughout the paper, concerning operators on the
bosonic Fock space. In Appendix \ref{sec:bounds}, we summarize 
bounds used to control the interaction between the electron (or
the nucleus) and the radiation field.

\section{The Model}\label{sec:model}

We consider a non-relativistic {\it atom} consisting of a {\it
nucleus} and an {\it electron} interacting through a two-body
potential $V(x)$. The Hamiltonian describing the dynamics of the
atom is the self-adjoint operator
\begin{equation}
H_{\text{atom}} = \frac{p_n^2}{2m_n} + \frac{p_e^2}{2m_e} + V (x_n
-x_e)\,
\end{equation}
acting on the Hilbert space $\H_{\text{atom}} = L^2 (\R^3, \rd
x_n) \otimes L^2 (\R^3, \rd x_e)$, where $x_n$ and $x_e$ denote
the position of the nucleus and of the electron, respectively, and
$p_n = -i \nabla_{x_n}$ and $p_e = -i \nabla_{x_e}$ are the
corresponding momenta. We assume that the interaction potential
$V(x)$ satisfies the following assumptions.
\begin{quote}
{\bf Hypothesis (H0):} The potential $V$ is a locally square
integrable function, with $\lim_{|x| \to \infty} V(x) = 0$, and
such that $V(x)$ is infinitesimally small with respect to the Laplace
operator $ p^2 = -\Delta$, in the sense that, for all $\eps
>0$ there exists a finite constant $C_{\eps} >0$ such that
\begin{equation}\label{eq:hypH0}
\| V \psi \| \leq \eps \| p^2 \psi \| + C_{\eps} \| \psi \| \,
,
\end{equation}
for every $\psi \in D(p^2) = H^2 (\R^3)$. Moreover the potential
$V$ is such that the ground state energy of the Schr\"odinger
operator $p^2 /2m + V$ (with $m^{-1} = m_e^{-1} + m_n^{-1}$) is
non-degenerate.
\end{quote}

\emph{Remarks.}
\begin{itemize}
\item[1)] Hypothesis (H0) is satisfied by the 
Coulomb potential $V(x) = -1/|x|$ and it is inspired by this potential.
It follows from (H0) that the Hamiltonian $\Hatom$ with domain $ H^2
(\R^3_{x_n} \times \R^3_{x_e})$ is a self-adjoint operator
on the Hilbert space $L^2 (\R^3 , \rd x_n) \otimes L^2 (\R^3 , \rd
x_e)$ and that it is bounded from below.
\item[2)] With little more effort we could have covered a much larger class of
locally square integrable potentials $V$ where only the negative part
$V_{-}(x)=\max\{-V(x),0\}$ is infinitesimally small with respect to
$\Delta$ and $\Hatom$ is self-adjointly realized in terms of a Friedrich's
extension. This would allow, e.g., for confining potentials that tend to
$\infty$ as $|x|\to\infty$.  
\item[3)] Note that we are neglecting the degrees of freedom corresponding
to the spin of the nucleus and of the electron, because they do
not play an interesting role in the scattering process.
\end{itemize}

Next we couple the atom to a quantized scalar radiation field. We
call the particles described by the quantized field {\it (scalar)
photons}. The pure states of the photon field are vectors in the
bosonic Fock space over the one-particle space $L^2 (\R^3, \rd
k)$,
\begin{equation}
\F = \bigoplus_{n \geq 0} L^2_s (\R^{3n} , \rd k_1 \dots \rd k_n)
\end{equation}
where $L^2_s (\R^{3n})$ denotes the subspace of $L^2 (\R^{3n})$
consisting of all functions which are completely symmetric under
permutations of the $n$ arguments. The variables $k_1, \dots k_n$
denote the momenta of the photons.

The dispersion relation of the photons is given by $\omega (k) =
|k|$, which characterizes relativistic particles with zero mass. The
free Hamiltonian of the quantized radiation field is given by the
second quantization of $\omega (k) =|k|$, denoted by $\dGamma
(|k|)$. Formally,
\begin{equation}
\dGamma (|k|) = \int \rd k \, |k| \, a^* (k) a (k) \, ,
\end{equation}
where $a^*(k)$ and $a (k)$ are the usual creation- and
annihilation operators on $\F$, satisfying the canonical
commutation relations $[a^{\sharp} (k) , a^{\sharp} (k') ] =0$,
$[a (k) , a^* (k')] = \delta (k - k')$. More notations for operators on Fock space that are used throughout the paper are collected in Appendix~\ref{sec:Fock}.

The total system, atom plus quantized radiation field, has the
Hilbert space $\H = \H_{\text{atom}} \otimes \F$; its dynamics is
generated by the Hamiltonian
\begin{equation}\label{eq:ham}
H_g = \Hatom + \dGamma (|k|) + g \left( \phi (G^e_{x_e}) + \phi
(G^n_{x_n})\right)
\end{equation}
where $g$ is a real non-negative coupling constant (the assumption
$g \geq 0$ is not needed, it just makes the notation a little bit
simpler), and where
\begin{equation}
\phi (G_{x}) = \int \rd k \, \left( a^* (k)  G_{x} (k) + a (k)
\overline{G}_{x} (k) \right).
\end{equation}
The form factors $G^e_{x}$ and $G_{x}^n$ are square integrable
functions of $k$ with values in the multiplication operators on
$L^2 (\R^3, \rd x)$. Clearly, $G_x^e$ describes the interaction
between the electron and the radiation field, and $G_x^n$ couples
the field to the nucleus. The next hypothesis specifies our
assumptions on the form factors $G^e_x$ and $G^n_x$.
\begin{quote}
{\bf Hypothesis (H1):} The form factors $G^e_x$ and $G^n_x$ have
the form
\begin{equation}\label{eq:Gx}
G^e_x (k) = e^{-ik\cdot x} \kappa_e (k) \quad \text{and} \quad
G^n_x (k) = e^{-ik\cdot x} \kappa_n (k)\, ,
\end{equation}
where $\kappa_e, \kappa_n$ belong to Schwartz space ${\mathcal S} (\R^3)$, 
and $\kappa_e(k) = \kappa_n (k) = 0$ if $|k| \leq \sigma$, for some $\sigma
>0$.
\end{quote}

The particular form of $G^e_x$ and $G^n_x$ given in (\ref{eq:Gx})
guarantees the translation invariance of the system (see the
discussion after (\ref{eq:trans})). The presence of an infrared
cutoff $\sigma >0$ in $\kappa_e$ and $\kappa_n$ is used in the
proofs of many of our results; but it is not necessary for the
existence of the asymptotic field operators and for the existence
of the wave operator in Section \ref{sec:waveop}. Notice that,
even though our main results require the coupling constant $g$ to
be sufficiently small, how small $g$ has to be does not depend on
the infrared cutoff $\sigma$, in the following sense. If we define
$\kappa_e (k) = \tilde{\kappa}_e (k) \chi (|k|/\sigma)$ and
$\kappa_n (k) = \tilde{\kappa}_n (k) \chi (|k|/\sigma)$, with
$\tilde{\kappa}_{e,n} \in C_0^{\infty} (\R^3)$, 
and with $\chi \in C^{\infty} (\R)$ monotone increasing and
such that $\chi (s) =0$ if $s \leq 1$ and $\chi (s) =1$ if $s \geq
2$, then Hypothesis (H1) is satisfied for every choice of $\sigma
>0$. Moreover how small $g$ has to be is independent of the choice
of the parameter $\sigma$.

Assuming Hypotheses (H0) and (H1), the Hamiltonian $H_g$, defined
on the domain $H^2 (\R^3_{x_e}\times \R^3_{x_n}) \otimes D
(\dGamma (|k|))$, is essentially self-adjoint and bounded from
below. This follows from Lemma~\ref{lm:estim}, which shows, using
Hypothesis (H1), that the interaction $\phi (G^e_{x_e}) + \phi
(G^n_{x_n})$ is infinitesimal with respect to the free Hamiltonian
$H_0 = \Hatom + \dGamma (|k|)$.

To study the system described by the Hamiltonian $H_g$, it is more
convenient to use coordinates describing the center of mass of the
atom and the relative position of the nucleus and the electron. We
define
\begin{equation}
X = \frac{m_n x_n + m_e x_e}{m_n + m_e} \, , \quad \quad x = x_e
-x_n \, .
\end{equation}
Then, the atomic Hamiltonian $\Hatom$ becomes
\begin{equation}
\Hatom = \frac{P^2}{2M} + \frac{p^2}{2m} + V(x)
\end{equation}
where $P = -i \nabla_X$ is the center of mass momentum of the
atom, and $p =- i\nabla_x$ is the momentum conjugate to the
relative coordinate $x$. Moreover, $M = m_e + m_n$ is the total
atomic mass, and $m =(m_e^{-1} + m_n^{-1})^{-1}$ is the reduced
mass. Expressed in the new coordinates, the total Hamiltonian of
the system is given by
\begin{equation}
\begin{split}
H_g &= \frac{P^2}{2M} + \frac{p^2}{2m} + V(x) +\dGamma (|k|) + g
\left( \phi (G^e_{X + \lambda_e x} ) + \phi (G^n_{X - \lambda_n
x}) \right) \\ &= \frac{P^2}{2M} + \frac{p^2}{2m} + V(x) +\dGamma
(|k|) + g \phi (G_{X,x}).
\end{split}
\end{equation}
Here $\lambda_e = m_n /M$ and $\lambda_n = m_e/M$, and we use the
notation
\begin{equation}\label{eq:GXx}
G_{X,x} (k) = G^e_{X+\lambda_e x} (k) + G^n_{X-\lambda_n x} (k) =
e^{-ik\cdot X} F_x (k),
\end{equation}
with
\begin{equation}\label{eq:Fx}
F_x (k) =  e^{-i\lambda_e k \cdot x} \kappa_e (k) + e^{i\lambda_n
k \cdot x} \kappa_n (k) \, .
\end{equation}
The fact that the form factors $\kappa_e$ and $\kappa_n$ contain
an infrared cutoff (meaning that $\kappa_n (k) = \kappa_e (k) =
0$, if $|k| \leq \sigma$) implies that photons with very small
momenta do not interact with the atom. In other words, they
decouple from the rest of the system. We denote by $\chi_i$ (the
subscript $i$ stands for ``interacting'') the characteristic
function of the set $\{ k \in \R^3:|k| \geq \sigma \}$. Then the
operator $\Gamma (\chi_i)$, whose action on the $n$-particle
sector of $\F$ is given by
\begin{equation}
\Gamma (\chi_i) = \chi_i \otimes \chi_i \otimes \dots \otimes
\chi_i,
\end{equation}
defines the orthogonal projection onto states without soft bosons.
The fact that soft bosons do not interact with the atom implies
that $H_g$ leaves the range of $\Gamma (\chi_i)$ invariant; $H_g$
commutes with $\Gamma (\chi_i)$. Another way to isolate the soft,
non-interacting, photons from the rest of the system is as
follows. We have that $L^2 (\R^3) = L^2 ( B_{\sigma} (0)) \oplus
L^2 (B_{\sigma} (0)^{c})$, where $B_{\sigma} (0)$ is the open ball
of radius $\sigma$ around the origin and $B_{\sigma} (0)^c$
denotes its complement. Hence the Fock space can be decomposed as
$\F \simeq\F_i \otimes \F_s$, where $\F_i$ is the bosonic Fock
space over $L^2 (B_{\sigma} (0)^c)$ (describing interacting
photons), and $\F_s$ is the bosonic Fock space over $L^2
(B_{\sigma} (0))$ (describing soft, non-interacting, photons).
Accordingly, the Hilbert space $\H= L^2 (\R^3, \rd X) \otimes L^2
(\R^3 , \rd x)\otimes \F$ can be decomposed as
\begin{equation} \begin{split}
\H &\simeq \H_i \otimes \F_s \quad \quad \text{with} \\
\H_i &= L^2 (\R^3 , \rd X) \otimes L^2 (\R^3 , \rd x) \otimes \F_i
.
\end{split}
\end{equation}
By $U:\H \to \H_i \otimes \F_s$ we denote the unitary map from
$\H$ to $\H_i \otimes \F_s$. The action of the Hamiltonian $H_g$
on $\H_i \otimes \F_s$ is then given by
\begin{equation}
U H_g U^* = H_i \otimes 1 + 1 \otimes \dGamma (|k|),
\end{equation}
with 
\begin{equation} 
H_i = H_g\restricted\H_{i} \, .
\end{equation}
Note that in the representation of the system on the Hilbert space
$\H_i \otimes \F_s$, the projection $\Gamma (\chi_i)$ (projecting
on states without soft bosons) is simply given by $U\Gamma
(\chi_i) U^* = 1 \otimes P_{\Omega}$, where $P_{\Omega}$ denotes
the orthogonal projection onto the vacuum $\Omega$ in $\F_s$.

One of the most important properties of the Hamiltonian $H_g$ is
its invariance with respect to translations of the whole system,
atom and field. More precisely, defining the total momentum of the
system by
\begin{equation}\label{eq:trans}
\Pi = P + \dGamma (k),
\end{equation}
we have that $[H_g, \Pi] = 0$. Because of this property, it can be
useful to rewrite the Hilbert space $\H = L^2 (\R^3 , \rd X)
\otimes L^2 (\R^3 , \rd x) \otimes \F$ as a direct integral over
fibers with fixed total momentum. Specifically, we define the
isomorphism
\begin{equation}
T: L^2 (\R^3, \rd X) \otimes L^2 (\R^3 , \rd x) \otimes \F
\longrightarrow L^2 (\R^3, \rd \Pi ; L^2 (\R^3 , \rd x) \otimes
\F)
\end{equation}
as follows. For $\psi = \{ \psi^{(n)} (X,x,k_1, \dots ,k_n) \}_{n
\geq 0} \in L^2 (\R^3 , \rd X)\otimes L^2 (\R^3,  \rd x) \otimes
\F$, we define
\begin{equation}
(T\psi)(\Pi) = \{ (T\psi)^{(n)}_{\Pi} (x, k_1 , \dots k_n) \}_{n
\geq 0} \in L^2 (\R^3 , \rd x) \otimes \F
\end{equation}
with \begin{equation}
 (T\psi)^{(n)}_{\Pi} (x, k_1 , \dots k_n) =
\widehat{\psi^{(n)}} (\Pi - k_1 - \dots -k_n, x, k_1, \dots k_n),
\end{equation}
where
\begin{equation}
\widehat{\psi^{(n)}} (P , x, k_1 ,\dots , k_n) =
\frac{1}{(2\pi)^{3/2}} \int \rd X e^{-iP\cdot X} \psi^{(n)}
(X,x,k_1,\dots, k_n)
\end{equation}
is the Fourier transform of $\psi^{(n)}$ with respect to its first
variable. Because of its translation invariance, the Hamiltonian
$H_g$ leaves invariant each fiber with fixed total momentum of the
Hilbert space $L^2 (\R^3, \rd \Pi; L^2 (\R^3 , \rd x) \otimes \F)
\simeq \int^{\oplus} (L^2 (\R^3 , \rd x) \otimes \F )\rd \Pi$.
More precisely,
\begin{equation}
\begin{split}
(T^* H_g T \psi) (\Pi) &= H_g(\Pi) \psi (\Pi) \quad \text{with }\\
H_g(\Pi)
&= \frac{(\Pi - \dGamma (k))^2}{2M} + \dGamma (|k|) + \Hato + g
\phi (F_x)
\end{split}
\end{equation}
where we put $\Hato = p^2 /2m + V$. Recall that $F_x (k) =
e^{-i\lambda_e k \cdot x} \kappa_e (k) + e^{i\lambda_n k \cdot x}
\kappa_n (k)$.

Note that, for every fixed $\Pi$, the operator $H_g(\Pi)$ is a
self-adjoint operator on the fiber space $L^2 (\R^3 , \rd
x)\otimes \F$. Our first results, stated in the next section,
describe the structure of the spectrum of the fiber-Hamiltonian
$H_g(\Pi)$, for fixed values of $\Pi$.

\section{The Spectrum of $H_g(\Pi)$}
\label{sec:spectrum}

\subsection{Dressed Atom States}\label{sec:das}

The first question arising in the analysis of the spectrum of
\begin{equation}
H_g(\Pi) = \frac{(\Pi - \dGamma (k))^2}{2M} + \dGamma (|k|) +
\Hato + g \phi (F_x)
\end{equation}
concerns the existence of a ground state of $H_g(\Pi)$: we wish to
know whether or not $E_g (\Pi) = \inf \sigma (H_g(\Pi))$ is an
eigenvalue of $H_g(\Pi)$. We will answer this question
affirmatively, under the assumption that the energy $E_g (\Pi)$
lies below some threshold and that the coupling constant is
sufficiently small. The restriction to small energies is necessary
to guarantee that the center of mass of the atom does not move
faster than with the speed of light ($c=1$ in our units), and that
the atom is not ionized. The following lemma (and its corollary)
proves that an upper bound on the total energy is sufficient to
bound the momentum of the center of mass of the atom (provided the coupling
constant is sufficiently small) and to make
sure that the electron is exponentially localized near the nucleus.

\begin{lemma}\label{lm:Sigma}
Assume that Hypotheses (H0) and (H1) are satisfied.
\begin{itemize}
\item[i)] Define $E^{\text{at}}_0 = \inf \sigma (\Hato)$, and fix
$\beta>0$. Suppose $\Sigma < \Sigma_{\beta} = E_0^{\text{at}} +
(M/2) \beta^2$. Then there is $g_{\Sigma,\beta}>0$ such that
\begin{equation}\label{eq:Sigma1}
\Big\| \frac{|\Pi - \dGamma (k)|}{M} \, E_{\Sigma} (H_g (\Pi))
\Big\| \leq \beta
\end{equation}
for all $g \leq g_{\Sigma,\beta}\ (g\geq 0)$, and for all $\Pi \in
\R^3$. In particular $\| (|P|/M) E_{\Sigma} (H_g) \| \leq \beta$.
\item[ii)] Define the ionization threshold
\[ \Sigma_{\text{ion}} = \lim_{R \to \infty} \inf_{\ph \in
\mathcal{D}_R} \langle \ph, H_g \ph \rangle
\] with $\mathcal{D}_R = \{ \ph \in D(H_g): \chi (|x| \geq R) \ph
= \ph \}$. Let $\Sigma,\alpha\in \R$ be such that
$\Sigma+\alpha^2/(2m)<\Sigma_{\text{ion}}$. Then
\begin{equation}\label{eq:Sigma2}
\sup_{\Pi\in \R^3}\| e^{\alpha |x|} E_{\Sigma} (H_g (\Pi)) \|
<\infty.
\end{equation}
\end{itemize}
\end{lemma}
\begin{proof}
i) Fix $\eps >0$ such that $\Sigma +\eps < \Sigma_{\beta}
=E^{\text{at}}_0 + (M/2)\beta^2$. Choose $\chi \in C^{\infty}_0
(\R)$ such that $\chi (s) =1$ for $s \leq \Sigma$, and $\chi
(s)=0$ if $s
>\Sigma +\eps$. Then we have that
\begin{equation}\label{eq:unif}
\begin{split}
\| (|\Pi - \dGamma (k)|/M) E_{\Sigma} (H_g (\Pi)) \| &\leq \|
(|\Pi - \dGamma (k)|/M) \chi (H_g (\Pi)) \| \\ &\leq \| (|\Pi -
\dGamma (k)|/M) \chi (H_0 (\Pi)) \| + C g
\end{split}
\end{equation}
where $H_0 (\Pi) = (\Pi - \dGamma (k))^2/2M + \Hato + \dGamma
(|k|)$ is the non-interacting fiber Hamiltonian and where the
constant $C$ is independent of $\Pi$. To prove the last equation
note that, if $\widetilde \chi$ denotes an almost analytic
extension of $\chi$ (see Appendix A in \cite{FGS3} for a short
introduction to the Helffer-Sj\"ostrand functional calculus), we
have that
\begin{equation}
\chi (H_g(\Pi)) - \chi (H_0 (\Pi)) = \frac{1}{\pi} \int \rd x \rd
y \, \partial_{\bar{z}} \widetilde \chi (z) \, \frac{1}{H_0 (\Pi)
-z} g \phi (F_x) \frac{1}{H_g (\Pi) -z}
\end{equation}
and therefore \begin{multline} \| |\Pi - \dGamma (k)| \, (\chi
(H_g(\Pi)) - \chi (H_0 (\Pi))) \| \\ \leq C g \| |\Pi - \dGamma
(k)| (H_0 (\Pi) +i)^{-1} \| \, \| \phi (F_x) (H_g(\Pi) +i)^{-1}\|
\leq C g \, ,
\end{multline}
uniformly in $\Pi$. Next, since $\Hato \geq E^{\text{at}}_0 = \inf
\sigma (\Hato)$, $\dGamma (|k|) \geq 0$, and by the definition of
$\chi$, we have that
\[ \chi (H_0 (\Pi)) = E_{\Sigma +\eps - E_0^{\text{at}}} \left(
\frac{(\Pi -\dGamma (k))^2}{2M} \right) \chi (H_0 (\Pi)) \, .\]
Since $\Sigma + \eps -E_0^{\text{at}} < (1/2)M\beta^2$, we
conclude from (\ref{eq:unif}) that
\[ \Big \| \frac{|\Pi - \dGamma (k)|}{M} \, E_{\Sigma} (H_g (\Pi)) \Big\|
 \leq \beta \] for $g$ sufficiently small
(independently of $\Pi$).

As for part ii), we use Theorem~1 of \cite{G} and an estimate from
its proof. Given $R\geq 0$ and $\Pi \in \R^3$, let
\begin{align*}
   \Sigma_{R} &= \inf_{\ph\in
   \mathcal{D}_{R},\|\ph\|=1}\sprod{\ph}{H_g\ph}\\
   \Sigma_{R} (\Pi) &= \inf_{\ph\in
   \mathcal{D}_{R,\Pi},\|\ph\|=1}\sprod{\ph}{H_g(\Pi)\ph}
\end{align*}
where $\mathcal{D}_{R} = \{ \ph \in D(H_g) : \chi (|x| \leq R) \ph
= 0 \}$ and $\mathcal{D}_{R,\Pi} = \{ \ph \in D(H_g (\Pi)) : \chi
(|x| \leq R) \ph = 0 \}$. Suppose for a moment that
\begin{equation}\label{eq:uni-pi}
  \Sigma_{R}(\Pi) \geq \Sigma_{R}
\end{equation}
for all $\Pi\in \R^3$ and all $R\in \R$. Then $\lim_{R\to\infty}
\Sigma_R (\Pi)\geq \Sigma_{\text{ion}}$ and hence
\[
   \|e^{\alpha|x|} E_{\Sigma}(H_g (\Pi))\| <\infty
\]
by \cite[Theorem~1]{G}. Moreover, the value of the parameter $R$
in the proof of \cite[Theorem~1]{G}, in the case of the
Hamiltonian $H_g(\Pi)$, can be chosen independent of $\Pi$ thanks
to \eqref{eq:uni-pi}. It follows that the estimate for
$\|e^{\alpha|x|} E_{\Sigma}(H_g(\Pi))\|$ from that proof is also
independent of $\Pi$. It thus remains to prove \eqref{eq:uni-pi}.
To this end we proceed by contradiction, assuming that
(\ref{eq:uni-pi}) is wrong. Then there exist $\Pi_0 \in \R^3$ and
$\eps >0$ such that
\[ \Sigma_R (\Pi_0) = \Sigma_R -\eps \,.\] Hence, we find $\ph_0
\in \mathcal{D}_{R,\Pi_0} \subset L^2 (\R^3 , \rd x) \otimes \F$
with $\| \ph_0 \| =1$ and with
\begin{equation}\label{eq:ph0Hph0}
\langle \ph_0 , H_g(\Pi_0) \ph_0 \rangle \leq \Sigma_R
-\frac{\eps}{2} \,.
\end{equation}
Moreover, since the map $\Pi \to \langle \ph_0, H_g (\Pi) \ph_0
\rangle$, for fixed $\ph_0$, is continuous in $\Pi$ (it is just a
quadratic function in $\Pi$), there exists $\delta >0$ such that
\[ \langle \ph_0, H_g (\Pi) \ph_0 \rangle \leq \Sigma_R -
\frac{\eps}{4} \] for all $\Pi$ with $|\Pi - \Pi_0| \leq \delta$.
Next, we choose $f \in L^2 (B_{\delta} (\Pi_0))$ (where
$B_{\delta} (\Pi_0)$ denotes the ball of radius $\delta$ around
$\Pi_0$) with $\| f \| =1$, and we define $\ph \in L^2 (\R^3, \rd
\Pi; L^2 (\R^3, \rd x) \otimes \F)$ by
\[ \ph (\Pi) = f(\Pi) \ph_0 \, .\] {F}rom (\ref{eq:ph0Hph0}), we
obtain that \begin{equation} \langle \ph, H_g \ph \rangle = \int
\rd \Pi \, \langle \ph (\Pi), H_g (\Pi) \ph (\Pi) \rangle\leq
(\Sigma_R - \eps/4),
\end{equation}
because $\| f \|=1$. Since $\ph \in \mathcal{D}_R$ (which is clear
from the construction of $\ph$), this contradicts the definition
of $\Sigma_R$.
\end{proof}

In the next proposition we prove the existence of a simple ground
state for $H_g(\Pi)$, provided the energy is lower than a
threshold energy $\Sigma$ and the coupling constant is small
enough.

\begin{prop}\label{prop:gs}
Assume Hypotheses (H0) and (H1) are satisfied. Fix $\beta <1$ and
choose $\Sigma < \min (\Sigma_{\beta} , \Sigma_{\text{ion}})$ (see
Lemma \ref{lm:Sigma} for the definition of $\Sigma_{\beta}$ and
$\Sigma_{\text{ion}}$). Then, for $g$ sufficiently small
(depending on $\beta$ and $\Sigma$), $E_g (\Pi) = \inf \sigma
(H_g(\Pi))$ is a simple eigenvalue of $H_g(\Pi)$, provided that $E_g
(\Pi) \leq \Sigma$.
\end{prop}

\emph{Remark.} Since $E_g(\Pi)\leq E_0^{\rm at}+\Pi^2/2M$ it suffices that 
\(E_0^{\rm at}+\Pi^2/2M\leq \Sigma\) and that $g$ is small enough.

\begin{proof} The proof is very similar to the proof of Theorem 4
in \cite{FGS3}. For completeness we repeat the main ideas, but we
omit details. In order to prove the proposition, we consider the
modified Hamiltonian $\Hmod$ (defined in Section \ref{sec:modham})
given, on the fiber with fixed total momentum $\Pi$, by
\begin{equation}
\Hmod (\Pi) = \frac{(\Pi - \dGamma (k))^2}{2M} + \dGamma (\omega)
+ \frac{p^2}{2m} +V(x) +g \phi (F_x),
\end{equation}
where the dispersion law $\omega (k)$ (with $\omega (k) = |k|$ if $|k| \geq
\sigma$ and $\omega (k) \geq \sigma/2$ for all $k$) 
is assumed to satisfy Hypothesis (H2) of Section
\ref{sec:modham}. Set $E_{\text{mod}} (\Pi) = \inf \sigma
(\Hmod(\Pi))$. Note that $\Hmod (\Pi)$ and $H_g(\Pi)$ act
identically on the range of $\Gamma (\chi_i)$, the orthogonal
projection onto the subspace of vectors without soft bosons.

The proof of the proposition is divided into four steps.
\begin{itemize}
\item[1)] Suppose $E_g (\Pi) \leq \Sigma$. Then, for sufficiently
small $g$ (depending on $\beta$ and $\Sigma$), we have that
\begin{equation}\label{eq:part1}
\inf_{|k| \geq \eps} E_g (\Pi -k) + |k| - E_g (\Pi) > 0 \, ,
\end{equation}
for every $\eps >0$. This inequality follows by perturbation of
the free Hamiltonian (see Lemma 35 in \cite{FGS3} for details).
\item[2)] $E_g (\Pi) = E_{\text{mod}} (\Pi)$. Moreover, if $\psi$
is an eigenvector of $H_g (\Pi)$ (or of $\Hmod (\Pi)$)
corresponding to the eigenvalue $E_g (\Pi)$, then $\psi \in \ran
\Gamma (\chi_i)$. In particular, $\psi$ is an eigenvector of
$H_g(\Pi)$ corresponding to the eigenvalue $E_g (\Pi)$ if and only
if $\psi$ is an eigenvector of $\Hmod (\Pi)$ corresponding to the
eigenvalue $E_{\text{mod}} (\Pi) = E_g (\Pi)$.
\end{itemize}
In order to prove these statements note that the Hamiltonians
$H_g(\Pi)$ and $\Hmod (\Pi)$ act on the fiber space $\H_{\Pi} =
L^2 (\R^3) \otimes \F = L^2 (\R^3) \otimes \F_i \otimes \F_s$,
where $\F_s$ is the Fock space of the soft bosons, $\F_s =
\oplus_{n\geq 0} L^2_s \left( B_{\sigma} (0)^{\times n}; \rd k_1
\dots \rd k_n \right)$. Thus
\[ \H_{\Pi} \simeq \bigoplus_{n \geq 0} L^2_s \left( B_{\sigma}
(0)^{\times n}, \rd k_1 \dots \rd k_n; L^2 (\R^3) \otimes \F_i
\right) =: \bigoplus_{n \geq 0} \H_{\Pi}^{(n)} \,. \] The
restriction of $H_g (\Pi)$ to the subspace $\H_{\Pi}^{(n)}$ with
exactly $n$ soft bosons is given by \[(H_g (\Pi) \psi) (k_1, \dots
,k_n) = H_{\Pi} (k_1 , \dots ,k_n) \psi (k_1 , \dots ,k_n)\] with
\begin{equation}\label{eq:sof}
\begin{split}
H_{\Pi} (k_1 , \dots, k_n) &= \frac{(\Pi - \dGamma (k) -
\sum_{j=1}^n k_j )^2}{2M} + \dGamma (|k|) + \sum_{j=1}^n |k_j| +
\frac{p^2}{2m} + V(x) + g \phi (F_x) \\ &= H_g (\Pi - \sum_{j=1}^n
k_j) + \sum_{j=1}^n |k_j|
\geq E_g (\Pi - \sum_{j=1}^n k_j) + \Big| \sum_{j=1}^n k_j \Big| \\
&> E_g (\Pi) \quad \quad \text{if} \quad (k_1 , \dots k_n) \neq
(0, \dots 0) \; .
\end{split}
\end{equation}
In the last inequality we used the result of part (1). This proves
that
\begin{equation}
E_g (\Pi) = \inf \sigma (H_g (\Pi)) = \inf \sigma (H_g(\Pi)|_{L^2
(\R^3)\otimes \F_i}) = \inf \sigma (\Hmod (\Pi)|_{L^2
(\R^3)\otimes \F_i}) \geq E_{\text{mod}} (\Pi)\, .
\end{equation}
Since $H_g (\Pi) \leq \Hmod (\Pi)$, we conclude that $E_g (\Pi) =
E_{\text{mod}} (\Pi)$. Eq.~(\ref{eq:sof}) also proves that
eigenvectors of $H_g (\Pi)$ corresponding to the energy $E_g
(\Pi)$, if they exist, belong to the range of $\Gamma (\chi_i)$.
That the same is true for eigenvectors of $\Hmod (\Pi)$
corresponding to the energy $E_g (\Pi)$ follows from an inequality
for $\Hmod (\Pi)$ analogous to (\ref{eq:sof}).
\begin{itemize} \item[3)] If $E_g (\Pi) \leq \Sigma$, and for $g$
sufficiently small (depending on $\beta$ and $\Sigma$) we have
that
\begin{equation}
\Delta (\Pi) = \inf_{k} E_g (\Pi - k) + \omega (k) - E_g (\Pi)
> 0 \, .
\end{equation}
\end{itemize}
For $|k| > \sigma /4$, this follows from part (1) (because $\omega
(k) \geq |k|$), while for $|k| \leq \sigma /4$, this inequality
follows from $E_g (\Pi -k) + \omega (k) - E_g (\Pi) = E_g (\Pi -k
) +|k| - E_g (\Pi) + (\omega (k) -|k|) \geq \sigma /4$, by
(\ref{eq:part1}) and by construction of $\omega$ (see Section
\ref{sec:modham}).
\begin{itemize}
\item[4)]
\begin{equation}
\inf \sigma_{\text{ess}} (\Hmod (\Pi)) \geq \min \left( E_g (\Pi)
+ \Delta (\Pi) , \Sigma_{\text{mod}} (\Pi) \right) \, ,
\end{equation}
where $\Sigma_{\text{mod}} (\Pi)$ is defined like
$\Sigma_{\text{ion}}$ with $H_g$ replaced by $\Hmod (\Pi)$.
(Recall from 2) that $E_{\text{mod}} (\Pi) = E_g (\Pi)$).
\end{itemize}
The proof of part 4) is very similar to the proof of Lemma 36 in
\cite{FGS3} with a small modification at the beginning. We first
need to localize with respect to the relative coordinate $x$. That
is, we choose $J_0,J_{\infty} \in C^{\infty} (\R^3, [0,1])$ with
$J_0 (x) =1$ for $|x| \leq 1$, $J_0 (x) =0$ for $|x| \geq 2$ and
$J^2_0 + J^2_{\infty} =1$. Let $J_{\sharp,R} (x) = J_{\sharp}
(x/R)$. Then \begin{equation} \label{eq:zero} \Hmod (\Pi) =
J_{0,R} \Hmod (\Pi) J_{0,R} + J_{\infty,R} \Hmod (\Pi)
J_{\infty,R} + O (R^{-2}) \end{equation} as $R \to \infty$. As in
\cite{FGS3}, one shows that
\[J_{0,R} \Hmod (\Pi) J_{0,R} \geq J_{0,R}^2 (E_g (\Pi) + \Delta
(\Pi)) + K \] with $K$ relatively compact w.r.t. $\Hmod (\Pi)$,
while
\[J_{\infty,R} \Hmod (\Pi)
J_{\infty,R} \geq J_{\infty,R}^2 \Sigma_{\text{ion}} (\Pi) +o(1)
\] as $R \to \infty$ follows from the definition of
$\Sigma_{\text{mod}} (\Pi)$. Part 4) follows from these estimates
applied to the r.h.s. of (\ref{eq:zero}).

Along with 1) and 2), and since $\Sigma_{\text{mod}} (\Pi) \geq
\Sigma_{\text{ion}}$ for every $\Pi$ (see (\ref{eq:uni-pi}) and
its proof), this proves that $E_g (\Pi)$ is an eigenvalue of $H_g
(\Pi)$, provided that $E_g (\Pi) \leq \Sigma$ and $g$ is sufficiently
small (depending on $\Sigma$). The proof of the fact that $E_g
(\Pi)$ is a simple eigenvalue is given in Corollary
\ref{cor:spectrum}, below.
\end{proof}

From now on, for fixed $\beta <1$ and $\Pi$ such that $E_g (\Pi)
\leq \Sigma < \min (\Sigma_{\beta}, \Sigma_{\text{ion}})$, we
denote by $\psi_{\Pi}$ the unique (up to a phase) normalized
ground state vector of $H_g(\Pi)$. The vector $\psi_{\Pi}$ is
called a {\it dressed atom state} with fixed total momentum $\Pi$.
The space of dressed atom wave packets, \(\Hgs \subset \H\), is
defined by
\begin{equation*}
T \Hgs = \Big\{ \psi \in L^2 \Big(\{\Pi : E_g (\Pi) \leq \Sigma \}
; \, L^2 (\R^3, \rd x) \otimes \F \Big): \psi(\Pi)\in \langle
\psi_{\Pi} \rangle \Big\}
\end{equation*}
where $\langle \psi_{\Pi} \rangle$ is the one-dimensional space
spanned by the vector $\psi_{\Pi}$; $\Hgs$ is a closed linear
subspace left invariant by the Hamiltonian $H_g$. In fact, $H_g$
commutes with the orthogonal projection $\Pgs$, onto $\Hgs$. This
follows from \(( U \Pgs U^* \ph )(\Pi) = P_{\psi_{\Pi}} \ph
(\Pi)\).

\subsection{The Fermi Golden Rule}

From Hypothesis (H0), and from standard results in the
theory of Schr\"odinger operators (see \cite{RS4}), it follows that the
spectrum of
\[ \Hato=\frac{p^2}{2m} +V\] in the negative half-axis $(-\infty,0)$ is
discrete. We denote the negative eigenvalues of $\Hato$ by
$E_0^{\text{at}} < E_1^{\text{at}} < \dots < 0$. The eigenvalues
$E_j^{\text{at}}$ can accumulate at zero only. If $\Hato$ has no
eigenvalues (a possibility which is not excluded by our
assumptions), then our results (which only concern states of the
system for which the electron is bound to the nucleus) are
trivial. We denote the (finite) multiplicity of the eigenvalue
$E_j^{\text{at}}$ by $m_j$. For fixed $j \geq 0$, we denote by
$\ph_{j,\alpha}$, $\alpha =1,\dots m_j$, an orthonormal basis of
the eigenspace of $\Hato$ corresponding to the eigenvalue
$E_j^{\text{at}}$. By Hypothesis (H0),
the lowest eigenvalue, $E_0^{\text{at}}$, of
$\Hato$ is simple ($m_0 =1$). The unique (up to a phase) ground
state vector of $\Hato$ is denoted by $\ph_0$.

\begin{figure}
\begin{center}
\epsfig{file=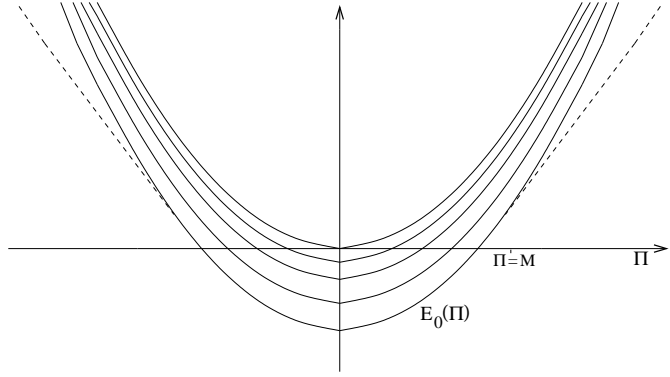,scale=.5}
\end{center}
\caption{Eigenvalues of the free Hamiltonian $H_0 (\Pi)$ (the
dashed curve represents $\inf \sigma (H_0
(\Pi))$).}\label{fig:spectrum}
\end{figure}

For every fixed $\Pi$, the free Hamiltonian,
\begin{equation}
H_0 (\Pi) =\frac{(\Pi - \dGamma (k))^2}{2M} + \dGamma (|k|) +
\Hato \, ,
\end{equation}
has eigenvalues \[E_j (\Pi) = E_j^{\text{at}} + \Pi^2/2M\] with
multiplicity $m_j$ corresponding to the eigenvectors $\psi_j =
\ph_{j,\alpha} \otimes \Omega$ (where $\Omega \in \F$ denotes the
Fock vacuum), for every $j \geq 0$. For $|\Pi| / M \leq 1$, $E_0
(\Pi) = \inf \sigma (H_0 (\Pi))$, while all other eigenvalues $E_j
(\Pi)$, $j \geq 1$, are embedded in the continuous spectrum
(see Fig. \ref{fig:spectrum}).
For $|\Pi|/M >1$, \emph{all} eigenvalues of $H_0 (\Pi)$ are embedded
in the continuous spectrum, and the Hamiltonian $H_0 (\Pi)$ does
not have a ground state. We restrict our attention to the
physically more interesting case $|\Pi|/M \leq 1$; (this will be
ensured by the condition that the total energy of the system is
less than $\Sigma_{\beta =1} = E_0^{\text{at}} + (1/2)M$ and that
the coupling constant $g$ is sufficiently small). One then expects
the embedded eigenvalues $E_j (\Pi)$, $j \geq 1$, to dissolve and
to turn into resonances when the perturbation $\phi (F_x)$ is
switched on. We prove that this is indeed the case, provided the
lifetime of the resonances, as predicted by Fermi's Golden Rule in
second order perturbation theory, is finite.

For given $i,j \geq 0$ and $k \in \R^3$, we define the $m_i \times
m_j$ matrix
\begin{equation}
(A_{ij} (k))_{\alpha, \alpha'} := \langle \ph_{i, \alpha}\, , F_x
(k) \ph_{j,\alpha'} \rangle \, .
\end{equation}
For given $j \geq 0$, we then define the resonance matrix by
setting
\begin{equation}
\Gamma_j (\Pi) = \sum_{i: i \leq j} \int \rd k \, A_{ji}^* (k)
A_{ji} (k) \, \delta \left( \frac{(\Pi -k)^2}{2M} + |k| -
\frac{\Pi^2}{2M} + E_i^{\text{at}} - E_j^{\text{at}} \right)\,,
\end{equation}
where $\delta (.)$ denotes the Dirac delta-function. Note that
$\Gamma_j (\Pi)$ is an $m_j \times m_j$ matrix. According to
second order perturbation theory (Fermi's Golden Rule), the $m_j$
eigenvalues of $\Gamma_j (\Pi)$ are inverses of the lifetimes of
the resonances bifurcating from the unperturbed eigenvalue $E_j
(\Pi) = E_j^{\text{at}} + \Pi^2 /2M$. We put
\begin{equation}
\gamma_j (\Pi) = \inf \sigma (\Gamma_j (\Pi)) \, .
\end{equation}
Instability of the eigenvalue $E_j (\Pi)$ is equivalent, in second
order perturbation theory, to the statement that $\gamma_{j} (\Pi)
>0$.
\begin{quote}
{\bf Hypothesis (H2):} For fixed $\beta <1$ and $\Sigma < \min (
\Sigma_{\beta}, \Sigma_{\text{ion}})$, we assume that
\begin{equation}\label{eq:H2}
\inf \{ \gamma_j (\Pi) : \Pi \in \R^3 , j \geq 1 \; \text{and} \;
E_j (\Pi) < \Sigma \} > 0 \, .
\end{equation}
\end{quote}

\subsection{The Positive Commutator}
\label{sec:poscomm}

In order to prove the absence of embedded eigenvalues we use the
technique of positive commutators. We prove the positivity of the
commutator between the Hamiltonian $H_g (\Pi)$ and a suitable
conjugate operator $A$. Then the absence of eigenvalues follows
with the help of a virial theorem. We make use of ideas from
\cite{BFSS}, adapting them to our problem.

For fixed $j \geq 1$, we construct a suitable conjugate operator
$A$, and, in Proposition \ref{prop:poscomm1}, we prove the
positivity of the commutator $[H_g(\Pi),iA]$ when restricted to an
energy interval $\Delta$ containing the unperturbed eigenvalue
$E_j (\Pi)$ but no other eigenvalues of $H_0 (\Pi)$. In
Proposition~\ref{prop:poscomm2}, we then establish a commutator estimate 
on an energy interval around the ground state
energy $E_0 (\Pi)$ of $H_0 (\Pi)$.

For fixed $j \geq 1$ we define
\begin{equation}
P_j = \sum_{\alpha} |\ph_{j,\alpha}\rangle \langle \ph_{j,\alpha}|
\otimes P_{\Omega}, \quad \quad P_{\Omega} =
|\Omega\rangle\langle\Omega|\,.
\end{equation}
By definition, $P_j$ is the orthogonal projection onto the
eigenspace of $H_0 (\Pi)$ corresponding to the eigenvalue $E_j
(\Pi) = \Pi^2 /2M + E_{j}^{\text{at}}$. We also define the
symmetric operator
\begin{equation}
A = \dGamma (a) +iD
\end{equation}
where
\begin{equation}\label{eq:a}
a = \frac{1}{2} \, \dGamma \left(\hat{k} \cdot y + y \cdot \hat{k}
\right), \qquad \text{with } \hat k = \frac{k}{|k|},
\end{equation}
and
\begin{equation}\label{eq:D}
D = g\theta P_j a (F_x) R_{\eps}^2
\overline{P}_j - g \theta \overline{P}_j R_{\eps}^2 a^* (F_x) P_j
\, .
\end{equation}
In (\ref{eq:D}), we introduced the notation $\overline{P}_j = 1 -
P_j$ and we used
\begin{equation}
R_{\eps}^2 = \left( (H_0 (\Pi) - E_j (\Pi))^2 + \eps^2
\right)^{-1}\,.
\end{equation}
Note that $\eps R_{\eps}^2 \to \delta (H_0 (\Pi) - E_j (\Pi))$ strongly,
as $\eps \to 0$. The real parameters $\theta$ and $\eps$ will be
fixed later on.
\begin{prop}\label{prop:poscomm1}
We assume that Hypotheses (H0)-(H2) are satisfied. We fix $\beta
<1$ and choose $\Sigma < \min (\Sigma_{\beta} ,
\Sigma_{\text{ion}})$; (see Lemma \ref{lm:Sigma} for the
definition of $\Sigma_{\beta}$ and $\Sigma_{\text{ion}}$).
Moreover, we suppose that the interval $\Delta \subset (-\infty ,
\Sigma)$ is such that $E_j(\Pi) \in \Delta$ and
\begin{equation}\label{eq:dist}
d := dist \left( \Delta , \sigma_{\text{pp}} (H_0 (\Pi))\backslash
\{ E_j (\Pi) \} \right) > 0 \, .
\end{equation}
Then, for $g >0$ sufficiently small (depending on $\beta$,
$\Sigma$ and the distance $d$), one can choose $\eps$ and $\theta$
such that
\begin{equation}
E_{\Delta} (H_g (\Pi)) [H_g(\Pi), iA] E_{\Delta} (H_g(\Pi)) \geq C
E_{\Delta} (H_g (\Pi)) \, ,
\end{equation}
with a positive constant $C$.
\end{prop}
{\it Remarks.}
\begin{enumerate}
\item[1)] The choice of the parameter $\eps,\theta$ and of the
constant $C$ depends on the value of $g$. We can choose, for
example, \[ \theta \simeq O (g^{\kappa}) \quad \quad \eps \simeq
O( g^{\alpha}) \quad \quad C \simeq O(g^{2+\kappa -\alpha}), \]
for $0< \kappa <\alpha <1$. \item[2)] How small $g$ has to be
chosen depends on the values of $\beta$, on $\gamma_j (\Pi)$, and
on the distance $d$ in (\ref{eq:dist}). We must have that $g \ll
(1-\beta)$, $\gamma_j (\Pi) \gg g^{\alpha -\kappa}$, and $\gamma_j
(\Pi) d^2 \gg \max (g^{\alpha -\kappa}, g^{1-\alpha})$ (for an
arbitrary choice of $\kappa,\alpha$ with $0 < \kappa < \alpha
<1$). Moreover, $g$ has to be sufficiently small, in order for Eq.~(\ref{eq:Sigma1}) 
to hold true (and thus $g$ depends on the choice
of the threshold $\Sigma$). \item[3)] In this proposition, we do
not need the infrared cutoff in the interaction (i.e., we can
choose $\sigma =0$). However, the infrared cutoff is needed in the
proof of Proposition~\ref{prop:virial} (the Virial Theorem), and
hence in the proof of the absence of embedded eigenvalues.
\end{enumerate}
\begin{proof}
We begin with a formal computation of the commutator $[H_g(\Pi),
iA]$:
\begin{equation*}
[H_g(\Pi),iA] = N - \frac{(\Pi -\dGamma (k))}{M} \cdot \dGamma
(\hat{k}) - g \phi (ia F_x) - [H_g(\Pi),D]\,.
\end{equation*}
Recall that $\hat k = k/|k|$. Using Eq. (\ref{eq:Sigma1}), it is
easy to check that, for $\Delta$ as above,
\begin{equation}
E_{\Delta} (H_g(\Pi)) \left( N - \frac{\Pi - \dGamma (k)}{M} \cdot
\dGamma (\hat k)\right) E_{\Delta}(H_g(\Pi)) \geq (1 -\beta)
E_{\Delta} (H_g(\Pi)) (1 -P_{\Omega}) E_{\Delta} (H_g(\Pi)) \, .
\end{equation}
Thus, defining
\begin{equation}
B = (1 -\beta) (1 -P_{\Omega}) - g \phi (ia F_x) - [H_g(\Pi), D]\,
,
\end{equation}
we conclude that
\begin{equation}\label{eq:commB}
E_{\Delta} (H_g(\Pi)) [ H_g(\Pi) , iA] E_{\Delta} (H_g(\Pi)) \geq
E_{\Delta} (H_g(\Pi)) B E_{\Delta} (H_g(\Pi)),
\end{equation}
and it is enough to prove the positivity of the r.h.s. of the last
equation to complete the proof. The advantage of working with $B$,
instead of the commutator $[H_g(\Pi), iA]$, is that $B$ is bounded
with respect to the Hamiltonian $H_g(\Pi)$ while $[H_g(\Pi),iA]$
is not (since the number operator $N$ is not bounded with respect
to $H_g(\Pi)$). In order to prove that
\begin{equation}\label{eq:pos1}
E_{\Delta} (H_g(\Pi)) B E_{\Delta} (H_g(\Pi)) \geq C E_{\Delta}
(H_g (\Pi)),
\end{equation} we first establish the inequality
\begin{equation}\label{eq:pos2}
E_{\Delta} (H_0 (\Pi)) B E_{\Delta} (H_0 (\Pi)) \geq C E_{\Delta}
(H_0 (\Pi)) \, .
\end{equation}
To this end we may assume that
\begin{equation}\label{eq:deflambda}
\lambda_0 := \inf \sigma \left( E_{\Delta} (H_0(\Pi)) B E_{\Delta}
(H_0(\Pi))|_{\ran E_{\Delta} (H_0 (\Pi))} \right) \leq
\frac{1-\beta}{2},
\end{equation}
for otherwise (\ref{eq:pos2}) holds with $C=(1-\beta)/2$. This
assumption will allow us to apply the Feshbach map with projection
$P_j$ to the operator $E_{\Delta} (H_0(\Pi)) B E_{\Delta}
(H_0(\Pi)) - \lambda_0$. Indeed, this operator restricted to $\ran
\overline{P}_j$ is invertible for small $g$ and $g^2 \theta
\eps^{-2}$ as is shown in Step 1. Step~1 through Step 5 prepare
the proof of (\ref{eq:pos2}).

\bigskip

\noindent {\it Step 1.} There exists a constant $C >0$,
independent of $\Pi$ and $g$, such that
\begin{equation}\label{eq:step1}
\overline{P}_j E_{\Delta} (H_0 (\Pi)) B E_{\Delta} (H_0 (\Pi))
\overline{P}_j \geq \left(1 - \beta - C \left( g + \frac{g^2
\theta}{\eps^2}\right)\right) \overline{P}_j E_{\Delta} (H_0
(\Pi)).
\end{equation}
In fact,
\begin{equation}
\begin{split}
\overline{P}_j E_{\Delta} (H_0 (\Pi)) B E_{\Delta} (H_0 (\Pi))
&\overline{P}_j \geq (1-\beta) \overline{P}_j E_{\Delta} (H_0
(\Pi)) (1 -P_{\Omega})E_{\Delta} (H_0 (\Pi)) \overline{P}_j \\ &+
g \overline{P}_j E_{\Delta} (H_0 (\Pi)) \phi (ia F_x) E_{\Delta}
(H_0 (\Pi)) \overline{P}_j \\ &- g^2 \theta \overline{P}_j
E_{\Delta} (H_0 (\Pi)) a^* (F_x) P_j a (F_x) R_{\eps}^2 
E_{\Delta} (H_0 (\Pi))\overline{P}_j
\\ &- g^2 \theta \overline{P}_j E_{\Delta} (H_0
(\Pi)) R_{\eps}^2 a^* (F_x) P_j a (F_x) E_{\Delta} (H_0
(\Pi))\overline{P}_j \, .
\end{split}
\end{equation}
Applying Lemma \ref{lm:phibounds} of Appendix \ref{sec:bounds} to
bound $\phi (ia F_x)$, using that $P_{\Omega} \overline{P}_j
E_{\Delta} (H_0 (\Pi)) = 0$ (by the choice of the interval
$\Delta$), and that $\| R_{\eps}^2 \| \leq \eps^{-2}$, inequality
(\ref{eq:step1}) follows easily.

\bigskip

\noindent {\it Step 2.} We define
\begin{equation}\label{eq:fesh}
\E = P_j B P_j - P_j B \overline{P}_j E_{\Delta} (H_0 (\Pi))
\left( (B -\lambda_0) |_{\ran \overline{P}_j E_{\Delta} (H_0
(\Pi))} \right)^{-1} E_{\Delta} (H_0 (\Pi)) \overline{P}_j B P_j\,
,
\end{equation}
where $\lambda_0$ is defined in (\ref{eq:deflambda}) (note that,
by (\ref{eq:deflambda}) and by the result of Step 1, the inverse
on the r.h.s. of (\ref{eq:fesh}) is well defined, if $g$ and $g^2
\theta/\eps^2$ are small enough). Then we have
\begin{equation}\label{eq:step2}
\lambda_0 \geq \inf \sigma \left( \E |_{\ran P_j} \right).
\end{equation}
The proof of (\ref{eq:step2}) relies on the isospectrality of the
Feshbach map and can be found, for example, in \cite{BFSS}. This
inequality says that, instead of finding a bound on the operator
$B$ restricted to the range of $E_{\Delta} (H_0 (\Pi))$, we can
study the operator $\E$ restricted to the much smaller range of
the projection $P_j$ (which is finite-dimensional).

Using the assumption (\ref{eq:deflambda}) and Eq.
(\ref{eq:step1}), we see that, if $g$ and $g^2 \theta /\eps^{2}$
are sufficiently small,
\begin{equation}\label{eq:condE}
(B- \lambda_0 )|_{\ran \overline{P}_j E_{\Delta} (H_0 (\Pi))} \geq
\frac{1-\beta}{4}.
\end{equation}
(Later, when we will choose the parameter $\theta$ and $\eps$, we
will make sure that $g^2 \theta/\eps^2$ is small enough, if $g$ is
small enough).  This implies that the operator $\E$ is bounded
from below by
\begin{equation}\label{eq:E}
\E \geq P_j B P_j - \frac{4}{1-\beta} P_j B \overline{P}_j
E_{\Delta} (H_0 (\Pi)) \overline{P}_j B P_j .
\end{equation}
Next, we study the second term on the r.h.s. of inequality
(\ref{eq:E}).

\bigskip

\noindent {\it Step 3.} There exists a constant $C >0$ independent
of $g,\theta$ and $\eps$ such that
\begin{equation}\label{eq:step3}
P_j B \overline{P}_j E_{\Delta} (H_0 (\Pi)) \overline{P}_j B P_j
\leq C g^2 P_j + C \left( g^2 \theta^2 + \frac{g^4
\theta^2}{\eps^2}\right) \, P_j a( F_x) R_{\eps}^2 a^* (F_x) P_j.
\end{equation}

In order to prove this bound, we note that, for any $\psi \in \H$,
\begin{equation}
\langle \psi, P_j B \overline{P}_j E_{\Delta} (H_0 (\Pi))
\overline{P}_j B P_j \psi \rangle = \| E_{\Delta} (H_0 (\Pi))
\overline{P}_j B P_j \psi \|^2 \, .
\end{equation}
Furthermore
\begin{equation}
\begin{split}
\overline{P}_j B P_j  = \; &-g \overline{P}_j \, \phi (i a F_x)
P_j  - \overline{P}_j \, [ H_g(\Pi) - E_j
(\Pi) , D] P_j  \\
= &- g \overline{P}_j \, \phi (iaF_x) P_j + g\theta \overline{P}_j
\, (H_g (\Pi) - E_j (\Pi)) \overline{P}_j \, R_{\eps}^2 a^* (F_x)
P_j \, ,
\end{split}
\end{equation}
because \[ P_j (H_g(\Pi) -E_j (\Pi)) P_j = P_j (H_0 (\Pi) - E_j
(\Pi)) P_j + g P_j \phi (F_x) P_j = 0 .\] Hence we find that
\begin{equation}
\begin{split}
E_{\Delta} (H_0 (\Pi)) \overline{P}_j B P_j \psi = \; &-g
E_{\Delta} (H_0 (\Pi)) \overline{P}_j \phi (i a F_x) P_j \psi \\
&+g\theta \overline{P}_j E_{\Delta} (H_0 (\Pi)) (H_0(\Pi) - E_j
(\Pi)) R_{\eps}^2 a^* (F_x) P_j \psi \\ &+ g^2 \theta
\overline{P}_j E_{\Delta} (H_0 (\Pi))\phi (F_x) \overline{P}_j
R_{\eps}^2 a^* (F_x) P_j \psi \, .
\end{split}
\end{equation}
Using Lemma \ref{lm:phibounds} and the bound $\| (H_0 (\Pi) - E_j
(\Pi)) R_{\eps} \| \leq 1$, we conclude that
\begin{equation}
\| E_{\Delta} (H_0 (\Pi)) \overline{P}_j B P_j \psi \| \leq C g \|
P_j \psi \| + C \left( g \theta + g^2 \theta /\eps\right) \|
R_{\eps} a^* (F_x) P_j \psi \|\,.
\end{equation}
Taking the square of this inequality, we obtain (\ref{eq:step3}).

\bigskip

\noindent {\it Step 4.} We show that
\begin{equation}\label{eq:step4}
P_j B P_j = 2 g^2 \theta P_j a (F_x) R_{\eps}^2 a^* (F_x) P_j \,.
\end{equation}
Using that $P_j (1-P_{\Omega}) P_j = 0$ and $P_j \phi (ia G_x) P_j
= 0$, we easily find that
\begin{equation}
P_j B P_j = g\theta P_j H_g(\Pi) \overline{P}_j R_{\eps}^2 a^*
(F_x) P_j + g\theta P_j a (F_x) R_{\eps}^2 \overline{P}_j H_g(\Pi)
P_j\,.
\end{equation}
Writing $H_g(\Pi) = H_0 (\Pi) + g \phi (F_x)$, using that $P_j$
commutes with $H_0 (\Pi)$, and that $P_j a^* (F_x) = a (F_x) P_j =
0$ we find
\begin{equation}
P_j B P_j = g\theta P_j a (F_x) \overline{P}_j R_{\eps}^2 a^*
(F_x) P_j + g\theta P_j a (F_x)
 R_{\eps}^2 \overline{P}_j a^* (F_x) P_j \, .
\end{equation}
Eq. (\ref{eq:step4}) now follows writing $\overline{P}_j = 1 -
P_j$ and using that $P_j a (F_x) P_j = P_j a^* (F_x) P_j =0$.

\bigskip

From Step 3 and Step 4 and from (\ref{eq:E}) we get
\begin{equation}\label{eq:E2}
\E \geq \left( 2 g^2 \theta - C \left( g^2 \theta^2 +
\frac{g^4\theta^2}{\eps^2}\right)\right) P_j a (F_x) R_{\eps}^2
a^* (F_x) P_j - C g^2 P_j
\end{equation}
for a constant $C$ independent of $g,\theta$ and $\eps$.

\bigskip

\noindent {\it Step 5.} Next, we claim that
\begin{equation}\label{eq:step5}
P_j a (F_x) R_{\eps}^2 a^* (F_x) P_j \geq \frac{\gamma_j
(\Pi)}{\eps} ( 1 + o_{\eps} (1)) P_j\, ,
\end{equation}
where $o_{\eps} (1) \to 0$, as $\eps \to 0$.

In order to prove \eqref{eq:step5}, we use the pull-through
formula for $a(q) R_{\eps}^2$ and the fact that $a(q)P_j =0$. This
yields
\begin{equation}
\begin{split}
P_j a (F_x) &R_{\eps}^2 a^* (F_x) P_j = \int \rd q \rd q' P_j
\overline{F_x (q)} a(q) R_{\eps}^2 F_x (q') a^* (q') P_j
\\ &= \int \rd q \rd q' P_j \overline{F_x (q)} \left(
\left(\frac{(\Pi - q - \dGamma (k))^2}{2M} + |q| +
\dGamma (|k|) + \Hato - E_j (\Pi)\right)^2 + \eps^2\right)^{-1} \\
& \hspace{2cm}
\times F_x (q') a (q) a^* (q') P_j \\
&= \int \rd q  \, P_j \overline{F_x (q)} \left( \left(\frac{(\Pi -
q - \dGamma (k))^2}{2M} + |q| + \dGamma (|k|) +
\Hato - E_j (\Pi)\right)^2 + \eps^2\right)^{-1} \\
&\hspace{2cm} \times F_x (q) P_j \,.
\end{split}
\end{equation}
Next, we write $P_j = P_j^{\text{at}} \otimes P_{\Omega}$, where
$P_j^{\text{at}} = \sum_{\alpha=1}^{m_j} |\phi_{j,\alpha}\rangle
\langle \phi_{j, \alpha} |$ is the orthogonal projection onto the
eigenspace of $\Hato$ corresponding to the eigenvalue $E_j^{\text{at}}$. Then
we obtain that
\begin{equation}\label{eq:PaR}
\begin{split}
P_j a &(F_x) R_{\eps}^2 a^* (F_x) P_j \\ &= \left\{ \int \rd q \,
P_j^{\text{at}} \overline{F_x (q)} \left( \left(\frac{(\Pi
-q)^2}{2M} +|q| -\frac{\Pi^2}{2M} + \Hato -
E_j^{\text{at}}\right)^2 + \eps^2\right)^{-1} F_x (q)
P_j^{\text{at}} \right\} \otimes P_{\Omega} .\end{split}
\end{equation}
All operators involved in the expression on the r.h.s. of
(\ref{eq:PaR}) act trivially on the Fock space, and we get the
lower bound
\begin{equation}
\begin{split}
\int \rd q \, P_j^{\text{at}} &\overline{F_x (q)}
\left(\left(\frac{(\Pi -q)^2}{2M} +|q| -\frac{\Pi^2}{2M} + \Hato -
E_j^{\text{at}}\right)^2 + \eps^2\right)^{-1} F_x (q)
P_j^{\text{at}} \\ &\geq \sum_{m \leq j} \rd q \, P_j^{\text{at}}
\overline{F_x (q)} P_m^{\text{at}} \left(\left(\frac{(\Pi
-q)^2}{2M} +|q| -\frac{\Pi^2}{2M} + \Hato -
E_j^{\text{at}}\right)^2 + \eps^2\right)^{-1} P_m^{\text{at}}
F_x (q) P_j^{\text{at}}\\
&= \sum_{m \leq j} \sum_{\alpha,\alpha'} \int \rd q \, \left(
\left(\frac{(\Pi -q)^2}{2M} + |q| - \frac{\Pi^2}{2M} +
E_m^{\text{at}} - E_j^{\text{at}}\right)^2 + \eps^2\right)^{-1} \\
&\hspace{2cm} \times  \left(A_{mj}^* (q) A_{mj}
(q)\right)_{\alpha,\alpha'} |\ph_{j,\alpha}\rangle \langle
\ph_{j,\alpha'}| \, .
\end{split}
\end{equation}
Using that $\eps (x^2 + \eps^2)^{-1} = \delta (x) + o_{\eps} (1)$,
as $\eps \to 0$, and recalling the definition of the matrix
$\Gamma_j (\Pi)$, we find that
\begin{equation}
\langle \psi , P_j a (F_x) R_{\eps}^2 a^* (F_x) P_j \psi \rangle
\geq \frac{1}{\eps} (1 + o_{\eps} (1)) \langle \psi, P_j \Gamma_j
(\Pi) P_j \psi \rangle \geq \frac{\gamma_j (\Pi)}{\eps} (1 +
o_{\eps} (1)) \| P_j \psi \|^2 \, .
\end{equation}
This proves Eq. (\ref{eq:step5}).

\bigskip

{\it Proof of Eq. \eqref{eq:pos2}. } {F}rom (\ref{eq:E2}) and
(\ref{eq:step5}), we derive that
\begin{equation}
\E \geq \left( \frac{\gamma_j (\Pi)}{\eps} \left( 2 g^2 \theta - C
\left( g^2 \theta^2 + \frac{g^4\theta^2}{\eps^2} \right)\right) (1
+ o_{\eps}(1)) - C g^2\right) P_j \, .
\end{equation}
Choosing $\eps = g^{\alpha}$ and $\theta = g^{\kappa}$, with $0 <
\kappa < \alpha <1$, we get
\begin{equation}
\E \geq \gamma_j (\Pi) g^{2+\kappa-\alpha} P_j \, ,
\end{equation}
for $g$ sufficiently small. Note that, with this choice of $\eps$
and $\theta$, $g^2 \theta /\eps^2 = g^{2+\kappa-2\alpha} \ll 1$,
and thus (\ref{eq:condE}) is satisfied, if $g$ is small enough.
From (\ref{eq:deflambda}) and (\ref{eq:step2}) we then get that
\begin{equation}\label{eq:H0}
E_{\Delta} (H_0 (\Pi)) B E_{\Delta} (H_0 (\Pi)) \geq \gamma_j
(\Pi) g^{2+\kappa-\alpha} E_{\Delta} (H_0 (\Pi))
\end{equation}
which proves (\ref{eq:pos2}).

\bigskip

{\it Proof of Eq. \eqref{eq:pos1}.} To prove (\ref{eq:pos1}), and
hence complete the proof of the proposition, we must replace
$E_{\Delta} (H_0 (\Pi))$ by the spectral projection $E_{\Delta}
(H_g (\Pi))$ of the full Hamiltonian $H_g(\Pi)$.

\bigskip

Given an interval $\Delta \subset (- \infty, \Sigma)$ with $E_j
\in \Delta$ and
\begin{equation}
\text{dist} \left( \Delta , \sigma_{\text{pp}} (H_0
(\Pi))\backslash \{ E_j \} \right) >0
\end{equation}
we choose an interval $\widetilde \Delta \subset (-\infty,
\Sigma)$ such that $\Delta \subset \widetilde \Delta$, and
\begin{equation}
\text{dist} \left( \widetilde \Delta , \sigma_{\text{pp}} (H_0
(\Pi)) \backslash \{ E_j \} \right) >0
\end{equation}
and with $\delta = \text{dist} (\Delta, \widetilde
\Delta^{\text{c}})
>0$. Furthermore, we choose a function $\chi \in C^{\infty} (\R)$
with the property that $\chi =1$ on $\Delta$ and $\chi =0$ on
$\widetilde \Delta^{\text{c}}$. We can assume that $|\chi' (s)|
\leq C \delta^{-1}$. Applying (\ref{eq:H0}) with $\Delta$ replaced
by $\widetilde \Delta$, and 
multiplying the resulting inequality with $\chi (H_0 (\Pi))$ we
get
\begin{equation}
\chi (H_0 (\Pi)) B \chi (H_0 (\Pi)) \geq \gamma_j (\Pi)
g^{2+\kappa-\alpha} \chi^2 (H_0 (\Pi)) \, .
\end{equation}
Setting $\chi := \chi (H_g (\Pi))$ and $\chi_0 := \chi (H_0
(\Pi))$, we have that
\begin{equation}
\chi B \chi = \chi_0 B \chi_0 + (\chi -\chi_0) B \chi_0 + \chi_0 B
(\chi -\chi_0) + (\chi -\chi_0) B (\chi -\chi_0) \, .
\end{equation}
Using
\begin{equation}
(\chi -\chi_0) B \chi_0 + \chi_0 B (\chi -\chi_0) \geq -(1/2)
\chi_0 B \chi_0 - 2 (\chi - \chi_0) |B| (\chi -\chi_0)
\end{equation}
we find that
\begin{equation}\label{eq:chiBchi}
\chi B \chi \geq 1/2 \chi_0 B \chi_0 - 3 (\chi -\chi_0) |B| (\chi
-\chi_0) \, .
\end{equation}
Next we use that
\begin{equation}
\begin{split}
\chi - \chi_0 = \; &\int \rd z \, \partial_{\bar{z}}
\widetilde{\chi} (z) \left( \frac{1}{H_g(\Pi) -z} - \frac{1}{H_0
(\Pi) -z}\right)  \\ = \; & g \int \rd z \,\partial_{\bar{z}}
\widetilde{\chi} (z) \frac{1}{H_g(\Pi) -z} \phi (F_x) \frac{1}{H_0
(\Pi) -z}\,.
\end{split}
\end{equation}
{F}rom the definition of $B$ it follows that \[ \| (H_0 (\Pi)
-z_1)^{-1} |B| (H_g(\Pi) -z)^{-1} \| \leq C (1 + g\theta
\eps^{-2}) = C (1 + g^{1+\kappa-2\alpha}) \, , \] with the choice
$\eps =g^{\alpha}$, $\theta = g^{\kappa}$. This implies that
\begin{equation}
\langle \psi , (\chi -\chi_0) |B| (\chi -\chi_0)\psi \rangle \leq
C \left( g^2 + g^{3+\kappa -2\alpha} \right) \| \psi\|^2 \, ,
\end{equation}
where the constant $C$ depends on $\delta$ ($C$ is proportional to
$\delta^{-2}$). Thus, with (\ref{eq:chiBchi}),
\begin{equation}
\begin{split}
\chi B \chi &\geq (1/2) \gamma_j (\Pi) g^{2+\kappa -\alpha}
\chi^2_0 - C (g^2 +g^{3+\kappa -2\alpha})  \\
&\geq (1/2) \gamma_j (\Pi) g^{2+\kappa-\alpha} \chi^2  - C (g^2
+g^{3+\kappa -2\alpha}) \, .
\end{split}
\end{equation}
Since $0 < \kappa < \alpha <1$, we have $g^{2+\kappa -\alpha} \gg
g^2$ and $g^{2+\kappa -\alpha} \gg g^{3+\kappa-2\alpha}$.
Therefore, multiplying from the left and the right with
$E_{\Delta} (H_g(\Pi))$, and choosing $g$ small enough, we find
that
\begin{equation}
E_{\Delta} (H_g (\Pi)) B E_{\Delta} (H_g(\Pi)) \geq C E_{\Delta}
(H_g(\Pi))\, ,
\end{equation}
for some positive constant $C$, which, with (\ref{eq:commB}),
completes the proof of the proposition.
\end{proof}

Proposition~\ref{prop:poscomm1} and Proposition~\ref{prop:virial},
below, prove absence of embedded eigenvalues of $H_g (\Pi)$ on
$(-\infty, \Sigma)$ with the exception of a small interval around the ground
state energy, $\inf \sigma (H_0 (\Pi))$. Absence of embedded
eigenvalues near the ground state energy follows from our next
proposition.

Recall from (\ref{eq:a}) the notation
\[ a :=\frac{1}{2} \,  \left( \hat k \cdot y + y \cdot \hat k
\right) , \] with $\hat k = k/|k|$.

\begin{prop}\label{prop:poscomm2}
Assume Hypotheses (H0)-(H1). Fix $\beta <1$, and choose $\Sigma <
\min (\Sigma_{\beta},\Sigma_{\text{ion}})$, with $\Sigma_{\beta} =
E_0^{\text{at}} + (M/2)\beta^2$. Suppose the interval $\Delta
\subset (-\infty, \Sigma)$ is such that
\begin{equation}
\Delta \subset (-\infty, E_1 (\Pi)) \quad \text{and } \quad d=
dist \, (\Delta , E_1 (\Pi)) >0 \, .
\end{equation}
(Recall that $E_1 (\Pi)$ denotes the first excited eigenvalue of
the free Hamiltonian $H_0 (\Pi)$). Then, if $g \geq 0$ is
sufficiently small (depending on $\beta$, $\Sigma$ and $d$), there
exists $C>0$ such that
\begin{multline}
E_{\Delta} (H_g(\Pi)) [ H_g(\Pi) , i \dGamma (a) ] E_{\Delta}
(H_g(\Pi))
\\ \geq (1- \beta) E_{\Delta} (H_g(\Pi)) \left(1 - P_{\ph_0 \otimes
\Omega} \right) E_{\Delta} (H_g(\Pi)) - C g E_{\Delta}
(H_g(\Pi))\, .
\end{multline}
\end{prop}

\begin{proof}
A simple calculation shows that
\begin{equation}
[H_g(\Pi) , i\dGamma (a) ] = N -\frac{\Pi - \dGamma (k)}{M} \cdot
\dGamma (\hat k) + g \phi (ia F_x) \, .
\end{equation}
Hence
\begin{equation}\label{eq:com}
\begin{split}
E_{\Delta} (H_g(\Pi)) &[H_g(\Pi), i\dGamma(a)] E_{\Delta} (H_g(\Pi)) \\
&\geq (1 -\beta) E_{\Delta} (H_g(\Pi)) N E_{\Delta} (H_g(\Pi)) + g
E_{\Delta}
(H_g(\Pi)) \phi (ia F_x) E_{\Delta} (H_g(\Pi)) \\
&\geq (1 - \beta - C g) E_{\Delta} (H_g(\Pi)) - (1- \beta)
E_{\Delta} (H_g(\Pi)) P_{\Omega} E_{\Delta} (H_g(\Pi)) \, .
\end{split}
\end{equation}
Here we use that, by Hypothesis (H1), $\| E_{\Delta} (H_g(\Pi))
|\Pi - \dGamma (k)|/M \| \leq \beta$, and that, by
Lemma~\ref{lm:phibounds}, $\| \phi (ia F_x) E_{\Delta}
(H_g(\Pi))\| \leq C$.

Next, we note that
\begin{equation}\label{eq:pomega}
\begin{split}
E_{\Delta} (H_g(\Pi)) P_{\Omega} E_{\Delta} (H_g(\Pi)) = \;
&E_{\Delta} (H_g(\Pi)) \left( \chi (H_{\text{at}} =
E_0^{\text{at}})\otimes P_{\Omega} \right) E_{\Delta} (H_g(\Pi)) \\
&+ E_{\Delta} (H_g(\Pi)) \left( \chi (H_{\text{at}} \geq
E_1^{\text{at}})\otimes P_{\Omega} \right) E_{\Delta} (H_g(\Pi)) \\
= \; &E_{\Delta} (H_g(\Pi)) P_{\ph_0 \otimes \Omega} E_{\Delta}
(H_g(\Pi)) \\ &+ E_{\Delta} (H_g(\Pi)) \chi (H_0 (\Pi) \geq E_1
(\Pi)) E_{\Delta} (H_g (\Pi))
\end{split}
\end{equation}
where $\ph_0$ is the unique (up to a phase) ground state vector of
the atomic Hamiltonian $\Hato$. Next we choose a function $\chi
\in C^{\infty} (\R)$ with $\chi (s) =0$ for $s \leq E_1 (\Pi) - d$
and $\chi (s) =1$ for $s \geq E_1 (\Pi)$. Eq. (\ref{eq:pomega})
then implies that
\begin{equation}\label{eq:pomega2}
\begin{split}
E_{\Delta} (H_g(\Pi)) &P_{\Omega} E_{\Delta} (H_g(\Pi)) \\ &\leq
E_{\Delta} (H_g(\Pi)) P_{\ph_0 \otimes \Omega} E_{\Delta}
(H_g(\Pi)) + E_{\Delta} (H_g(\Pi)) \chi (H_0 (\Pi)) E_{\Delta}
(H_g (\Pi))\, .
\end{split}
\end{equation}
Note that
\begin{equation}
\begin{split}
\chi (H_0 (\Pi)) - \chi (H_g(\Pi)) &= \int \rd z
\partial_{\bar{z}} \widetilde \chi (z) \, \left( \frac{1}{H_0
(\Pi) - z} - \frac{1}{H_g (\Pi) -z} \right) \\ &= Cg \int \rd z
\partial_{\bar{z}} \widetilde \chi (z) \frac{1}{H_0 (\Pi) -z} \phi
(F_x) \frac{1}{H_g(\Pi) -z} \, .
\end{split}
\end{equation}
Since, by definition of the interval $\Delta$, $\chi (H_g (\Pi))
E_{\Delta} (H_g(\Pi)) = 0$, we find that
\begin{equation}
\begin{split}
E_{\Delta} (H_g(\Pi)) \chi (H_0 (\Pi)) E_{\Delta} (H_g (\Pi)) &=
E_{\Delta} (H_g(\Pi)) \left( \chi (H_g (\Pi)) - \chi (H_0 (\Pi))
\right) E_{\Delta} (H_g (\Pi)) \\ &\leq C g E_{\Delta}
(H_g(\Pi))\,.
\end{split}
\end{equation}
With (\ref{eq:com}) and (\ref{eq:pomega2}), this shows that
\begin{equation}
\begin{split}
E_{\Delta} (H_g(\Pi)) [H_g(\Pi), i\dGamma &(a)] E_{\Delta}
(H_g(\Pi))
\\ &\geq (1-\beta) E_{\Delta} (H_g(\Pi)) \left(1 - P_{\ph_0 \otimes
\Omega}\right) E_{\Delta} (H_g(\Pi)) - C g E_{\Delta} (H_g(\Pi))
\, .
\end{split}
\end{equation}
\end{proof}

\begin{prop}[Virial Theorem] \label{prop:virial}
Let Hypotheses (H0)-(H1) be satisfied. Assume that $H_g(\Pi) \ph =
E \ph$, for some $\ph \in L^2 (\R^3) \otimes \F$ with $\Gamma
(\chi_i) \ph = \ph$, and for an energy $E < \Sigma_{\text{ion}}$.
Then
\[ \sprod{\ph}{[H_g (\Pi) , i\dGamma (a)] \ph} =0 \quad \text{and}
\quad  \sprod{\ph}{[H_g(\Pi) , iA] \ph} =0,\] where $a$ is as
(\ref{eq:a}), and $A= \dGamma (a) + i D$ with $D$ defined in
(\ref{eq:D}).
\end{prop}

\begin{proof} To prove the proposition, we replace the Hamiltonian $H_g(\Pi)$ by
a modified Hamiltonian \[ \Hmod (\Pi) = \frac{(\Pi - \dGamma
(k))^2}{2M} + \frac{p^2}{2m} + V(x) + \dGamma (\omega) + g \phi
(F_x)\] where the new dispersion law $\omega (k) \in C^{\infty}
(\R^3)$ satisfies $\omega (k) = |k|$, for $|k| > \sigma$, and
$\omega (k) \geq \sigma /2$ for all $k$. Since the two Hamiltonian
$H_g(\Pi)$ and $\Hmod (\Pi)$ act identically on states without
soft bosons (in the range of the projection $\Gamma (\chi_i)$), it
is enough to prove that
\begin{equation}\label{eq:vir1}
\sprod{\ph}{[\Hmod (\Pi), i\dGamma (a_{\text{mod}})]\ph} =0
\end{equation}
and
\begin{equation}\label{eq:vir2}
\sprod{\ph}{[\Hmod (\Pi), iA_{\text{mod}}]\ph} = 0
\end{equation}
where $a_{\text{mod}} = (1/2) \dGamma (\nabla \omega (k) \cdot y +
y \cdot \nabla \omega (k))$, and $A_{\text{mod}}$ equals $A$ with
$a$ replaced by $a_{\text{mod}}$. The proof of (\ref{eq:vir1}) is
very similar to the proof of Lemma 40, in \cite{FGS3}. The proof
of (\ref{eq:vir2}) follows easily from (\ref{eq:vir1}), because
$A_{\text{mod}} - \dGamma (a_{\text{mod}})$ is a bounded operator
on $L^2 (\R^3, \rd x) \otimes \F$.
\end{proof}

\begin{corollary}\label{cor:spectrum}
Assume Hypotheses (H0)-(H2). Fix $\beta <1$ and choose $\Sigma <
\min (\Sigma_{\beta}, \Sigma_{\text{ion}})$, with $\Sigma_{\beta}$
and $\Sigma_{\text{ion}}$ as in Lemma \ref{lm:Sigma}. Then, for
sufficiently small values of $g
>0$,
\begin{equation}
\sigma_{\text{pp}} (H_g(\Pi)) \cap (-\infty,\Sigma) = \{ E_g
(\Pi)\},
\end{equation}
where $E_g (\Pi)$ is a simple eigenvalue, for all $\Pi$ with $E_g
(\Pi) \leq \Sigma$.
\end{corollary}

{\it Remark.} How small $g$ has to be chosen depends on the choice
of $\beta$ ($g\ll 1- \beta$), on the choice of $\Sigma$ (we need
(\ref{eq:Sigma1}) to hold true), on $\gamma_j (\Pi)$ ($g \ll \inf
\{ \gamma_j (\Pi) : j \geq 1, E_g (\Pi) \leq \Sigma \}$), and it
also depends on the distances between the eigenvalues of the
atomic Hamiltonian (we must require that $g^{1/2} \ll \min \{
|E_{j+1}^{\text{at}} - E_j^{\text{at}}| : 0\leq j \leq n\}$, where
$n$ is such that $E_n^{\text{at}} \leq \Sigma <
E_{n+1}^{\text{at}}$).

\begin{proof}
We first prove that if $g$ is sufficiently small, then
\begin{equation}\label{eq:cor1}
\sigma_{\text{pp}} \left( H_g(\Pi) \restricted{\ran \Gamma
(\chi_i)}\right) \cap (-\infty,\Sigma) = \{ E_g (\Pi) \}\, ,
\end{equation}
and that $E_g (\Pi)$ is a simple eigenvalue of
$H_g(\Pi)\restricted{\ran \Gamma (\chi_i)}$.

To this end, we define $\Delta_0 = (-\infty,(E_1 (\Pi) + E_0
(\Pi))/2)$. We define intervals \[ \Delta_j = \left( \frac{(E_j
(\Pi) + 2 E_{j-1} (\Pi))}{3} , \frac{(E_{j} (\Pi) + E_{j+1}
(\Pi))}{2}\right) \, , \] for $j= 1, \dots n$, where $n$ is such
that $E_{n-1} (\Pi) < \Sigma \leq E_n (\Pi)$. Each interval
$\Delta_j$ contains exactly one eigenvalue of the free Hamiltonian
$H_0 (\Pi)$, and $(-\infty,\Sigma) \subset \cup_{j=0}^n \Delta_j$.
The absence of eigenvalues of $H_g(\Pi)\restricted{\ran \Gamma
(\chi_i)}$ inside $\Delta_j$, for $j \geq 1$ and for $g$
sufficiently small, follows from Propositions \ref{prop:poscomm1}
and \ref{prop:virial}. Next, suppose that $\psi$ is a normalized
eigenvector of $H_g(\Pi)$ corresponding to an eigenvalue $E \in
\Delta_0$. Without loss of generality, we can assume that $\langle
\psi , \ph_0 \otimes \Omega \rangle$ is real; recall that $\ph_0$
is the unique (up to a phase) normalized ground state vector of
$\Hato=p^2/2m + V(x)$. Then, by Proposition \ref{prop:poscomm2}
and Proposition \ref{prop:virial}, we have that
\begin{equation}\begin{split}
0 &\geq (1 -\beta) \langle \psi, (1 - P_{\ph_0 \otimes \Omega})
\psi \rangle - C g = (1-\beta) \left( 1 - \langle \psi,\ph_0
\otimes \Omega\rangle^2\right) - C g \\ &\geq (1- \beta) \left(1 -
\langle \psi , \ph_0 \otimes \Omega\rangle \right) - C g =
\frac{1-\beta}{2} \| \psi - \ph_0 \otimes \Omega\|^2 - C g.
\end{split}
\end{equation}
Hence
\begin{equation}\label{eq:normpsi}
\| \psi - \ph_0 \otimes \Omega \|^2 \leq \frac{2 C g}{1-\beta}\, .
\end{equation}
If there were two orthogonal eigenvectors of
$H_g(\Pi)\restricted{\ran \Gamma (\chi_i)}$, $\psi_1$ and
$\psi_2$, corresponding to eigenvalues in $\Delta_0$ then both
would satisfy inequality (\ref{eq:normpsi}), and, thus, we would
conclude that
\begin{equation}
\|\psi_1 - \psi_2 \| \leq 2\sqrt{\frac{2 C g}{1-\beta}}\, .
\end{equation}
But this is impossible if $g \leq (1-\beta)/4C$. So, for $g$ small
enough, there can only be one eigenvector of
$H_g(\Pi)\restricted{\ran \Gamma (\chi_i)}$ corresponding to an
eigenvalue in $\Delta_0$. In Proposition \ref{prop:gs}, we have
proven that $H_g(\Pi)\restricted{\ran \Gamma (\chi_i)}$ has a
ground state vector. This proves the fact that $E_g (\Pi)$ is a
simple eigenvalue of $H_g(\Pi)\restricted{\ran \Gamma (\chi_i)}$
as well as the fact that $H_g(\Pi)\restricted{\ran\Gamma
(\chi_i)}$ has no other eigenvalue in $\Delta_0$. Hence
(\ref{eq:cor1}) follows. To complete the proof of the corollary,
we need to show that
\begin{equation}
\sigma_{\text{pp}} \left(H_g (\Pi)\restricted{(\ran \Gamma
(\chi_i))^{\perp}} \right)= \emptyset \, .
\end{equation}
To this end, we decompose $\F \simeq \F_i \otimes \F_s \simeq
\oplus_{n \geq 0} L^2_s (B_{\sigma} (0)^{\times n}, \rd k_1 \dots
\rd k_n ; \F_i)$, and we write $L^2 (\R^3 , \rd x) \otimes \F
\simeq \oplus_{n \geq 0} \H_n$, where \[ \H_n = L^2_s (B_{\sigma}
(0)^{\times n}, \rd k_1 \dots k_n; L^2 (\R^3, \rd x)\otimes \F_i)
\] is the space of vectors containing exactly $n$ soft,
non-interacting, bosons. The Hamiltonian leaves each $\H_n$
invariant, and the restriction of $H_g(\Pi)$ on $\H_n$ is given by
\begin{equation}
\begin{split}
(H_g(\Pi)\restricted{\H_n} \psi) (k_1,\dots k_n) &= H_{\Pi}
(k_1,\dots, k_n) \psi
(k_1, \dots k_n) \\
H_{\Pi} (k_1, \dots, k_n) &= H_g ( \Pi - k_1- \dots -k_n) +
\sum_{j=1}^n |k_j|\, .
\end{split}
\end{equation}
Here $H_g (\Pi - k_1 - \dots - k_n )$ is an operator over $L^2
(\R^3, \rd x) \otimes \F_i$, the space of states with no soft
bosons. We know that the only eigenvalue of $H_g (\Pi - k_1- \dots
- k_n)$ in $(-\infty, \Sigma)$ is its ground state energy $E_g
(\Pi - k_1 - \dots - k_n)$ as long as $E_g (\Pi - k_1 -\dots -k_n)
<\Sigma$. In particular the only eigenvalue of $H_{\Pi} (k_1,
\dots , k_n)$ in $(-\infty, \Sigma)$ is given by $E_g (\Pi -k_1
-\dots k_n) + |k_1|+\dots +|k_n|$ if this number is less than
$\Sigma$. Thus $E \in (-\infty, \Sigma)$ is an eigenvalue of
$H_g(\Pi)\restricted{\H_n}$ if and only if there exists a set $M
\subset B_{\sigma} (0)^{\times n}$with positive measure, such that
\begin{equation}
E = E_g (\Pi -k_1 - \dots - k_n) + |k_1| + \dots + |k_n|
\end{equation}
for all $(k_1, \dots ,k_n)\in M$. But this is impossible because
\[ |\nabla E_g (\Pi)| = |\sprod{\psi_{\Pi}}{(\Pi -\dGamma (k))/M
\psi_{\Pi}}| \leq 1\] for every $\Pi$ with $E_g (\Pi) < \Sigma$
and $g$ small enough.
\end{proof}

\section{Scattering Theory}
\label{sec:scatt}

The proofs of most of the results in this section are similar to
those of the corresponding results in \cite{FGS3}. In order to
give an idea of the structure of the proof of our main result
(asymptotic completeness, Theorem \ref{thm:ACph}), we repeat
here the most important theorems, but we omit most of their proofs (we
refer to the corresponding statements in \cite{FGS3}). The main
difference with respect to \cite{FGS3} is encountered in the proof
of the positivity of the asymptotic observable in Section
\ref{sec:pos}: there, we propose some new ideas to control the
internal degrees of freedom of the atom (which are not present in
\cite{FGS3}, because there we considered 
free electrons coupled to the quantized
radiation field).

\subsection{The Wave Operator}
\label{sec:waveop}

The first step towards understanding scattering theory for the
model studied in this paper consists in the construction of states
with asymptotically free photons. This can be accomplished using
asymptotic field operators, which are constructed in the next
theorem. Note that in Theorems \ref{thm:asy_fields} and
\ref{thm:Omega_+} we do not impose any infrared cutoff on the
interaction; we can take $\sigma =0$, provided the form factor
$\kappa (k)$ is smooth at $k=0$. We use the notation
$L^2_{\omega} (\R^3) = L^2 (\R^3 , (1 + 1/|k|) \rd k)$.

\begin{theorem}[Existence of asymptotic field operators]
\label{thm:asy_fields} Assume Hypotheses (H0)-(H1) are satisfied
(but $\sigma =0$ is allowed!). Fix $\beta <1$ and choose $\Sigma <
\min (\Sigma_{\beta},\Sigma_{\text{ion}})$ (with $\Sigma_{\beta}$
as in Lemma \ref{lm:Sigma}). If $g \geq 0$ is so small that
(\ref{eq:Sigma1}) is true, then the following results hold.
\begin{enumerate}
\item[i)] Let $h \in L^2_{\omega} (\R^3 )$ and let $h_t (k) =
e^{-i|k|t} h(k)$. Then the limit
\[a_+^{\sharp} (h) \ph = \lim_{t \to \infty} e^{iH_g t} a^{\sharp}
(h_t) e^{-iH_g t} \ph \] exists for all $\ph \in \ran E_{\Sigma}
(H_g)$. \item[ii)] Let $h,g \in L^2_{\omega} (\R^3 )$. Then
\[ [ a_+ (g) , a^*_+ (h)] = (g,h) \quad \text{and} \quad [
a^{\sharp}_+ (g) , a^{\sharp}_+ (h)] = 0,
\] in the sense of quadratic forms on $\ran E_{\Sigma} (H_g)$ ($a^{\sharp}
(h)$ means either $a^*(h)$ or $a(h)$).
\item[iii)] Let $h \in L^2_{\omega} (\R^3)$, and let $M:=\sup \{
|k| :h(k)\neq 0\}$ and $m := \inf \{ |k| :h(k)\neq 0\}$. Then
\begin{align*}
a_+^* (h) \ran \chi (H_g \leq E ) &\subset \ran \chi (H_g \leq E+M) \\
a_+ (h) \ran \chi (H_g \leq E ) &\subset \ran \chi (H_g \leq E-m),
\end{align*}
if $E\leq \Sigma$. \item[iv)] Let $h_i \in L^2_{\omega} (\R^3)$
for $i=1,\dots n$. Put $M_i = \sup \{ | k | :h_i (k)\neq 0\}$ and
assume $\ph \in \ran E_{\lambda} (H_g)$. Then if $\lambda +
\sum_{i=1}^n M_i \leq \Sigma$ we have that $\ph \in D(a_+^{\sharp}
(h_1) \dots a_+^{\sharp} (h_n))$, the limits
\[ a_+^{\sharp} (h_1) \dots a_+^{\sharp} (h_n)\ph = \lim_{t \to \infty}
e^{iH_g t} a^{\sharp} (h_{1,t}) \dots a^{\sharp} (h_{n,t})e^{-iH_g
t} \ph \] exist, and \[ \| a_+^{\sharp} (h_1) \dots a_+^{\sharp}
(h_n) (H_g +i)^{-n/2} \| \leq C \| h_1 \|_{\omega} \dots \| h_n
\|_{\omega} .
\]
\item[v)] If $\ph \in E_{\Sigma} (H_g) \Hgs$ and $h \in
L^2_{\omega} (\R^3)$,
\[ a_+ (h) \ph =0 .\]
(Wave packets of dressed atom states are vacua of the asymptotic
field operators).
\end{enumerate}
\end{theorem}

The proof of this theorem is very similar to the one of Theorem 13
and Lemma 14 in \cite{FGS3}. It relies on a propagation estimate
for the center of mass of the atom (see Proposition~12 in
\cite{FGS3}), which guarantees that if the energy is smaller than
$\Sigma_{\beta}$, then the asymptotic velocity of the atom is
bounded above by $\beta$ (here $\beta<1$), and it exploits the
fact that, because the energy is below the ionization threshold,
the electron is exponentially bound to the nucleus. These two
facts and the fact that the propagation speed of photons is the
speed of light are sufficient to prove that the interaction
between the atom and asymptotically freely propagating photons
tends to zero, as $t \to \infty$.

The existence of asymptotic field operators allows us to introduce
the wave operator $\Omega_+$ of the system. In order to define
$\Omega_+$, we add a new copy of the Fock space $\F$ describing
states of free photons to the physical Hilbert space $\H = L^2
(\R^3, \rd X) \otimes L^2 (\R^3, \rd x) \otimes \F$ . We define
the extended Hamiltonian
\begin{equation}
\Hex_g = H_g\otimes 1 + 1 \otimes \dGamma (|k|)
\end{equation}
on the extended Hilbert space $\widetilde{\H} = \H \otimes \F$. In
the next theorem, we establish the existence of the wave operator
$\Omega_+$ as an isometry from a subspace of $\widetilde \H$ to a
subspace of the physical Hilbert space $\H$. The ``scattering
identification map'', $I$, used in the definition of the wave
operator $\Omega_+$, is defined in Appendix \ref{sec:scattiden}.

\begin{theorem}[Existence of the wave operator]\label{thm:Omega_+}
Let Hypotheses (H0)-(H1) be satisfied (but $\sigma =0$ is
allowed). Fix $\beta <1$, and choose $\Sigma < \min
(\Sigma_{\beta} ,\Sigma_{\text{ion}})$ (with $\Sigma_{\beta},
\Sigma_{\text{ion}}$ defined as in Lemma \ref{lm:Sigma}). Then if
$g \geq 0$ is small enough (depending on $\beta$ and $\Sigma$) the
limit
\begin{equation}\label{eq:Omega_+}
  \Omega_+ \ph := \lim_{t \to \infty} e^{iH_g t}I e^{-i\Hex_g t}
  (\Pgs \otimes  1)\ph
\end{equation}
exists for an arbitrary vector $\ph$ in the dense subspace of
$\ran E_{\Sigma}(\Hex)$ spanned by finite linear combinations of
vectors of the form \(\gamma \otimes a^* (h_1) \dots a^* (h_n)
\Omega\), where $\gamma = E_{\lambda}(H_g)\gamma$, $h_i\in
L^2_{\omega}(\R^3)$, and with
\(\lambda+\sum_{i}\sup\{|k|:h_i(k)\neq 0 \}\leq \Sigma\). If
\(\ph=\gamma \otimes a^* (h_1) \ldots a^*(h_n)\Omega\) belongs to
this space then
\begin{equation}\label{eq:Omega_+2}
  \Omega_+ \ph = a_+^* (h_1) \dots a_+^* (h_n) \Pgs \gamma.
\end{equation}
Furthermore $\| \Omega_+ \| =1$, and $\Omega_+$ has therefore a
unique extension, also denoted by $\Omega_+$, to
$E_{\Sigma}(\Hex_g)\HxF$. On $(\Pgs \otimes 1)E_{\Sigma}(\Hex_g)\HxF$,
the operator $\Omega_+$ is isometric. For all $t\in \R$,
\[ e^{-iH_g t} \Omega_+ = \Omega_+ e^{-i\Hex_g t}.\]
\end{theorem}

For the proof of this theorem we refer to the proof of Theorem 15
in \cite{FGS3}, which is almost identical. From Eq.
(\ref{eq:Omega_+2}) we see that vectors in the range of $\Omega_+$
are limits of linear combination of vectors describing wave packet
of dressed atom states and configurations of finitely many
asymptotically freely moving photons. Physically, it is expected
that the asymptotic evolution of every state with an energy below
the ionization threshold of the atom (that is $\Sigma <
\Sigma_{\text{ion}}$) and so small that the atom does not
propagate with a velocity larger than one (i.e., $\Sigma <
\Sigma_{\beta=1}$) can be approximated by linear combinations of
vectors describing a dressed atom state and a configuration of
finitely many freely propagating photons. More precisely, one
expects that \[ \ran \Omega_+ \supset \ran E_{\Sigma} (H_g), \quad
\text{if} \quad \Sigma <\min (\Sigma_{\beta=1},
\Sigma_{\text{ion}}).\] This statement is called {\it asymptotic
completeness of Rayleigh scattering}. Due to technical
difficulties, we can only prove asymptotic completeness for states
with energy less than a threshold energy $\Sigma < \min
(\Sigma_{\beta=1/3}, \Sigma_{\text{exp}})$ and assuming that the
coupling constant $g$ is small enough. The following theorem is
our main result.

\begin{theorem}[Asymptotic Completeness]\label{thm:ACph} Assume
that Hypotheses (H0)-(H2) are satisfied (see Eqs.
(\ref{eq:hypH0}), (\ref{eq:Gx}), and (\ref{eq:H2})). Fix $\beta
<1/3$, and choose $\Sigma < \min (\Sigma_{\beta},
\Sigma_{\text{ion}})$ (with $\Sigma_{\beta}$ and
$\Sigma_{\text{ion}}$ as in Lemma \ref{lm:Sigma}). Then, for $g
> 0$ sufficiently small,
\[ \ran \Omega_+ \supset E_{\Sigma} (H_g) \H . \]
\end{theorem}
{\it Remark.} The allowed range of values of $g$ depends
on the value of $(1/3-\beta)$; (we need that $g \ll
1/3-\beta$), on the choice of $\Sigma$ ($g$ must be small enough
in order for Eq. (\ref{eq:Sigma1}) to hold true), on the value of
$\alpha = \inf \{ \gamma_j (\Pi): j \geq 1, E_g (\Pi) < \Sigma \}$
($g \ll \alpha$), and on $\delta= \min \{ |E_{j+1}^{\text{at}} -
E_j^{\text{at}}| : 0\leq j \leq n\}$, with $n$ such that
$E_n^{\text{at}} \leq \Sigma < E_{n+1}^{\text{at}}$ ($g^{1/2} \ll
\delta$). As remarked in Section \ref{sec:model}, the assumption
that $g$ is positive is not necessary, it only simplifies the
notation (but $g=0$ is not allowed, because in this case the fiber
Hamiltonian $H_g (\Pi)$ has embedded eigenvalues).

Theorem \ref{thm:ACph} will be seen to follow from Lemma
\ref{lm:AC}, where we show that it suffices to prove an analogous
statement for a modified Hamiltonian $\Hmod$ (introduced in the
next section) and from Theorem \ref{thm:AC} in Section
\ref{sec:pos}, where asymptotic completeness for $\Hmod$ is
proven.

\subsection{The Modified Hamiltonian}
\label{sec:modham}

The fact that the bosons are massless leads to some technical
difficulties connected with the unboundedness of the operator
$N=\dGamma(1)$ with respect to the Hamiltonian. However, as long
as the infrared cutoff is strictly positive, the number of bosons
with energy below $\sigma$ is conserved. This allows us to
introduce a modified Hamiltonian, where the dispersion law of the
soft, non-interacting, photons is changed. We define
\begin{equation*}
\Hmod = \frac{P^2}{2M} + \frac{p^2}{2m} + V(x) + \dGamma (\omega
)+ g \phi (G_{X,x}),
\end{equation*}
and we assume that the dispersion law $\omega$ has the following
properties.
\begin{quote}
{\bf Hypothesis (H3).} $\omega \in \mathcal{C}^{\infty} (\R^3)$,
with $\omega(k)\geq |k|$, $\omega (k)=|k|$, for $|k| \geq \sigma$,
$\omega (k)\geq \sigma /2$, for all $k \in \R^3$,
\(\sup_{k}|\nabla\omega(k)|\leq 1\), and $\nabla\omega(k)\neq 0$
unless $k=0$. Furthermore, \(\omega(k_1+k_2)\leq
\omega(k_1)+\omega(k_2)\) for all $k_1,k_2\in\R^3$. (Here $\sigma
>0$ is the infrared cutoff defined in Hypothesis (H1).)
\end{quote}

The two Hamiltonians, $H_g$ and $\Hmod$, agree on states
of the system without soft bosons. Recall that $\chi_i (k)$ is the
characteristic function of the set $\{k:|k| \geq \sigma \}$ and
that the operator $\Gamma (\chi_i)$ is the orthogonal projection
onto the subspace of vectors describing states without soft
bosons. It is straightforward to check that $H_g$ and $\Hmod$
leave the range of the projection $\Gamma (\chi_i)$ invariant and
that
\begin{equation}\label{eq:int_ham}
H_g \restricted{\ran \Gamma (\chi_i)} = \Hmod  \restricted{\ran
\Gamma (\chi_i)} .
\end{equation}
The same conclusion can be reached using the unitary operator \(U
: \H \to \H_i \otimes \F_s\) introduced in
Section~\ref{sec:model}. On the factorized Hilbert space \(\H_i
\otimes\F_s\), the Hamiltonians $H_g$ and $\Hmod$ are given by
\begin{equation}\label{eq:fact_ham}
\begin{split}
U H_g U^* &= H_i \otimes 1 + 1 \otimes \dGamma (|k|) \\
U \Hmod U^* &= H_i \otimes 1 + 1 \otimes \dGamma (\omega) \quad \text{with} \\
H_i &= \frac{P^2}{2M} + \frac{p^2}{2m} + V(x) + \dGamma (|k|) + g
\phi (G_{X,x}),
\end{split}
\end{equation}
and we see explicitly that the two Hamiltonians agree on states
without soft bosons.

The modified Hamiltonian $\Hmod$, just like the physical
Hamiltonian $H_g$, commutes with spatial translations, i.e.,
$[\Hmod , \Pi] =0$, where $\Pi = P + \dGamma (k)$ is the total
momentum of the system. In the representation of the system on the
Hilbert space $L^2 (\R^3_{\Pi}; L^2 (\R^3, \rd x) \otimes \F)$,
the modified Hamiltonian $\Hmod$ is given by
\begin{equation*}
\begin{split}
(T \Hmod T^* \psi )(\Pi) &= \Hmod (\Pi) \psi (\Pi), \\
\Hmod (\Pi) &= \frac{(\Pi- \dGamma (k))^2}{2M} + \frac{p^2}{2m} +
V(x) + \dGamma (\omega) + g \phi( F_x),
\end{split}
\end{equation*}
where $T: \H \to L^2 (\R^3 , d\Pi; L^2 (\R^3 , \rd x) \otimes \F)$
has been defined in Section \ref{sec:model}.

The fiber Hamiltonians $H_g (\Pi)$ and $\Hmod(\Pi)$ commute with
the projection $\Gamma (\chi_i)$ and agree on its range,
\begin{equation}\label{eq:restr}
H_g (\Pi)\restricted{\ran \Gamma (\chi_i)} = \Hmod (\Pi)
\restricted{\ran \Gamma (\chi_i)}.
\end{equation}

In the proof of Proposition \ref{prop:gs} we have shown that if
$\beta <1$ and $\Sigma < \min(\Sigma_{\beta},\Sigma_{\text{ion}})$ 
then, for $g$ small enough,
\begin{equation*}
\inf \sigma (\Hmod (\Pi)) = \inf \sigma (H_g (\Pi)) = E_g (\Pi),
\end{equation*}
where $E_{g}(\Pi)$ is a simple eigenvalue of $H_g (\Pi)$ and of
$\Hmod (\Pi)$, as long as $E_g (\Pi) \leq \Sigma$. Moreover, the
corresponding dressed atom states coincide. Since the subspace
$\Hgs$ is defined in terms of the dressed atom states
$\psi_{\Pi}$, it follows that vectors in $\Hgs$ also describe dressed
atom wave packets for the dynamics generated by the modified
Hamiltonian $\Hmod$.

We remark that
\[ \sigma_{\text{pp}} (\Hmod (\Pi)) \cap (-\infty
, \Sigma) = \{E_g (\Pi) \}, \] for all $\Pi \in \R^3$ with $E_g
(\Pi) \leq \Sigma$, and for $g$ sufficiently small; see Eq.
\eqref{eq:cor1} and Corollary~\ref{cor:spectrum}.

Next, we discuss the scattering theory for the modified
Hamiltonian $\Hmod$. As in Theorem~\ref{thm:Omega_+} we fix
$\beta<1$ and we choose $\Sigma < \min (\Sigma_{\beta}, \Sigma_{\text{ion}}
)$. Then,
by the assumption that $\omega(k)=|k|$ for $|k|\geq \sigma$ , and
since $\dGamma (|k|-\omega)$ commutes with $H_g$ and $\Hmod$ we
have that
\begin{equation}\label{eq:a_mod}
\begin{split}
e^{i\Hmod t} a^{\sharp} (e^{-i\omega t} h) e^{-i\Hmod t} &=
e^{iH_g t} e^{-i \dGamma (|k| - \omega) t} a^{\sharp} (e^{-i\omega
t} h) e^{i \dGamma (|k| - \omega) t}e^{-i H_g t} \\ &= e^{iH_g t}
a^{\sharp} (e^{-i|k|t} h) e^{-iH_g t}\, ,
\end{split}
\end{equation}
for all $t$. It follows that the limit
\begin{equation*}
a_{\text{mod},+}^{\sharp} (h) \ph =\lim_{t \to \infty} e^{i\Hmod
t} a^{\sharp} (e^{-i\omega t} h) e^{-i\Hmod t} \ph
\end{equation*}
exists and that $a^{\sharp}_{\text{mod},+} (h) \ph = a^{\sharp}_+
(h) \ph$, for all $\ph \in \ran E_{\Sigma} (\Hmod) \subset \ran
E_{\Sigma} (H_g)$ and for all $h \in L^2_{\omega} (\R^3)$. This
and the fact that vectors in $\Hgs$ describe dressed atom states
for $H_g$ and for $\Hmod$ show that the asymptotic states
constructed with the help of the Hamiltonians $H_g$ and $\Hmod$
coincide.

On the extended Hilbert space $\HxF =\H \otimes \F$, we define the
extended modified Hamiltonian \[ \Hmodex = \Hmod \otimes 1 + 1
\otimes \dGamma (\omega) . \] In terms of $\Hmod$ and $\Hmodex$ we
also define a modified version, $\tilde{\Omega}_+^{\text{mod}}$,
of the wave operator $\Omega_+$ introduced in Section
\ref{sec:waveop}.

\begin{lemma}\label{lm:tilde-Omega-mod}
Let Hypotheses (H0), (H1) and (H3) be satisfied ($\sigma =0$ in
Hypothesis (H1) is allowed, and then $\Hmod = H_g$). Fix $\beta
<1$ and $\Sigma < \min (\Sigma_{\beta}, 0)$. Then if $g \geq 0$ is
sufficiently small, depending on $\beta$ and $\Sigma$, the limit
\begin{equation}\label{eq:tildeOmega}
  \tilde{\Omega}_+^{\text{mod}} \ph = \lim_{t \to \infty} e^{i\Hmod
  t} I e^{-i\Hmodex t} \ph
\end{equation}
exists for all \(\ph \in E_{\Sigma} (\Hmodex) \HxF\). Moreover,
the modified wave operator $\Omega_+^{\text{mod}}$, defined by
$\Omega_+^{\text{mod}} = \tilde{\Omega}_+^{\text{mod}}(\Pgs
\otimes 1)$, agrees with $\Omega_{+}$ on $\ran E_{\Sigma}(\Hmodex)$. That is,
\begin{equation}\label{eq:Omega_+mod}
   \Omega_+^{\text{mod}} \ph = \Omega_+ \ph,
\end{equation}
for all $\ph \in \ran E_{\Sigma}(\Hmodex)\subset \ran
E_{\Sigma}(\tilde{H_g})$.
\end{lemma}
We now extend the domain of $\Omega_{+}$ to include arbitrarily
many soft, non-interacting bosons. As a byproduct we obtain a
proof of \eqref{eq:Omega_+mod}. To start with, we recall the
isomorphism $U: \F \to \F_i \otimes \F_s$ introduced in Section
\ref{sec:model} and define a unitary isomorphism $U \otimes U:
\HxF \to \H_i \otimes \F_i \otimes \F_s \otimes \F_s$ separating
interacting from soft bosons in the extended Hilbert space $\HxF$.
With respect to this factorization the extended Hamiltonian $\Hex$
becomes $\Hex_g = \Hex_i \otimes 1 \otimes 1 + 1 \otimes 1 \otimes
\dGamma (|k|) \otimes 1 + 1 \otimes 1 \otimes 1 \otimes \dGamma
(|k|)$, where $\Hex_i = H_i \otimes 1 + 1 \otimes \dGamma (|k|)$.
As an operator from $\H_i \otimes \F_i \otimes \F_s \otimes \F_s$
to $\H_i \otimes \F_s$, the wave operator $\Omega_+$ acts as
\begin{equation}\label{eq:wo_fact}
U \Omega_+ (U^* \otimes U^*) = \Omega_+^{\text{int}} \otimes
\Omega_+^{\text{soft}}
\end{equation}
where \(\Omega_+^{\text{int}}: \H_i \otimes \F_i \to \H_i \) is
given by
\begin{equation}\label{eq:wo_fact2}
\Omega_+^{\text{int}} = s-\lim_{t \to \infty} e^{iH_i t}I
e^{-i\Hex_i t} (\Pgs^{\text{int}} \otimes 1)
\end{equation}
while \( \Omega_+^{\text{soft}} : \F_s \otimes \F_s \to \F_s \) is
given by
\begin{equation}\label{eq:Omega_soft}
\Omega_+^{\text{soft}} = I (P_{\Omega} \otimes 1) ,
\end{equation}
where $P_{\Omega}$ is the orthogonal projection onto the vacuum
vector $\Omega \in \F_s$. In view of \eqref{eq:wo_fact} and
\eqref{eq:wo_fact2}, the domain of $\Omega_{+}$ can obviously be
extended to $\ran E_{\Sigma} (\Hex_i) \otimes \F_s \otimes
\F_s\supset \ran E_{\Sigma} (\Hex_g)$. For the modified wave
operator $\Omega_+^{\text{mod}} = \tilde{\Omega}_+^{\text{mod}}
(\Pgs\otimes 1)$, we have $\Omega_+^{\text{mod}}=\Omega_{+,
\text{mod}}^{\text{int}} \otimes \Omega_+^{\text{soft}}$, and from
$H_g \restricted{\ran \Gamma (\chi_i )} = \Hmod \restricted{\ran
\Gamma (\chi_i )}$ it follows that $\Omega_{+,
\text{mod}}^{\text{int}}=\Omega_{+}^{\text{int}}$. Consequently,
also $\Omega_+^{\text{mod}}$ is well defined on $\ran E_{\Sigma}
(\Hex_i) \otimes \F_s \otimes \F_s$ and $\Omega_+^{\text{mod}} =
\Omega_+$.

We summarize the main conclusions in a lemma.

\begin{lemma}\label{lm:AC}
Let the assumptions of Lemma~\ref{lm:tilde-Omega-mod} be
satisfied, and let \(\Omega_{+}\) be defined on \(\ran
E_{\Sigma}\otimes\F_{s}\otimes\F_{s}\), as explained above. Then
\begin{equation}\label{eq:ran-Omega}
  \ran \Omega_+ \cong \ran \Omega_+^{int} \otimes \F_s
\end{equation}
in the factorization $\H\cong\H_i \otimes \F_s$. In particular,
the following statements are equivalent:
\begin{enumerate}
\item[i)] $\ran \Omega_+ \supset E_{\Sigma} (H_g) \H$. \item[ii)]
$\ran \Omega_+ \supset \Gamma (\chi_i) E_{\Sigma} (H_g) \H $.
\item[iii)] $\ran \Omega_+ \supset E_{\Sigma} (\Hmod) \H$.
\item[iv)] $\ran \Omega_+ \supset \Gamma (\chi_i) E_{\Sigma}
(\Hmod) \H$.
\end{enumerate}
\end{lemma}

\subsection{Existence of the Asymptotic Observable and of the Inverse Wave Operator}
\label{sec:observable}

Fix $\beta < 1$ and choose $\Sigma <\min (\Sigma_{\beta},
\Sigma_{\text{ion}})$ (recall from Lemma \ref{lm:Sigma} that
$\Sigma_{\beta} = E_0^{\text{at}} + M\beta^2/2$). We choose
numbers $\beta_1, \beta_2, \beta_3$ and $\gamma$ such that
\[  \beta < \beta_1 < \beta_2 < \beta_3 < \gamma \, . \]
\begin{quote}
{\bf Definition. } We pick a function \(\chi_{\gamma}\in
\C^{\infty}(\R;[0,1])\) such that $\chi_{\gamma}\equiv 1$ on
$[\gamma,\infty)$ and $\chi_{\gamma}\equiv 0$ on
$(-\infty,\beta_3]$. Our \emph{asymptotic observable} $W$ is
defined by
\[ W= s-\lim_{t\to\infty} e^{i\Hmod
t} f (\Hmod) \dGamma (\chi_{\gamma} (|y| /t)) f(\Hmod) e^{-i\Hmod
t} ,\] where the energy cutoff $f$ is smooth and supported in
$(-\infty,\Sigma)$. For the existence of this limit, see
Proposition \ref{prop:W} below.
\end{quote}
The physical meaning of the asymptotic observable is easy to
understand: $W$ measures the number of photons that are
propagating with an asymptotic velocity larger than $\gamma$. We
will prove in Section \ref{sec:pos} that $W$ is \emph{positive}
when restricted to the subspace of vectors orthogonal to the space
$\Hgs$ of wave packets of dressed atom states. Instead of
inverting the wave operator $\Omega_+$ directly, we can then
invert it with respect to the asymptotic observable $W$. More
precisely, we define an operator $W_+:\H \to \HxF=\H \otimes \F$,
called the \emph{inverse wave operator}, such that $W = \widetilde
\Omega_+ W_+$. Then, using the positivity of $W$, we can construct
an inverse of $\widetilde \Omega_+$. In order to define $W_+$, we
need to split each boson state into two parts, the second part
being mapped to the second Fock-space of prospective
asymptotically freely moving bosons.
\begin{quote}
{\bf Definition. } We define $j_t:\h = L^2 (\R^3,dk) \to \h \oplus
\h$ as follows: let \(j_th=(j_{0,t}h,j_{\infty,t}h)\), where
$j_{\sharp,t}(y)=j_{\sharp}(|y|/t)$, $j_{\sharp}\in
C^{\infty}(\R;[0,1])$, $j_0+j_{\infty}\equiv 1$, $j_0\equiv 1$ on
$(-\infty,\beta_2]$, $\supp(j_0)\subset(-\infty,\beta_3]$ while
$j_{\infty}\equiv 1$ on $[\beta_3,\infty)$ and
$\supp(j_{\infty})\subset[\beta_2,\infty)$. Then the inverse wave
operator $W_+$ is given by
\[ W_+ = s-\lim_{t\to\infty} e^{i\Hmodex t} f(\Hmodex)
\uGamma(j_t)\dGamma(\chi_{\gamma,t}) f (\Hmod) e^{-i\Hmod t} ,\]
where $f$ is a smooth energy cutoff supported in
$(-\infty,\Sigma)$. See Appendix \ref{sec:factfock} for the
definition of the operator $\uGamma (j_t)$. For the existence of
this limit, see Proposition~\ref{prop:W+}.
\end{quote}
Note that, since by definition $\beta <\gamma$, the photons which
propagate with velocity larger than $\gamma$ are asymptotically
free. To prove this fact, notice first that Lemma \ref{lm:Sigma}
continues to hold with $H_g$ replaced by $\Hmod$. Hence, the
assumption that $\supp f \subset (-\infty, \Sigma)$ (where $f$ is
the energy cutoff appearing in the definition of $W$ and $W_+$)
with $\Sigma < \min (\Sigma_{\beta} , \Sigma_{\text{ion}})$
guarantees, for sufficiently small $g$, that both the nucleus and
the electron remain inside a ball of radius $\beta t$ around the
origin. In fact, using the assumption $\Sigma < \Sigma_{\beta}$
(and $g$ small enough), we can prove, analogously to Proposition
12 in \cite{FGS3}, that
\begin{equation}\label{eq:PEX}
s-\lim_{t \to \infty} h (|X|/t) f(\Hmod) e^{-it\Hmod} = 0
\end{equation}
for any $h \in C^{\infty} (\R)$ with $h' \in C^{\infty}_0 (\R)$,
$\supp h \subset (\beta,\infty)$ and for any $f \in C^{\infty}_0
(\R)$ with $\supp f \subset (-\infty, \Sigma)$. Recall that $X$ is
the coordinate of the center of mass of the atom. Moreover, the
assumption that $\Sigma <\Sigma_{\text{ion}}$ implies that the
electron and the nucleus remain exponentially bound for all times;
therefore, both the electron and the nucleus are localized inside
the ball of radius $\beta t$. As a consequence, the interaction
strength between the nucleus (or the electron) and those bosons
counted by $\dGamma(\chi_{\gamma} (|y|/t))$ decays in $t$ at an
integrable rate. To establish this fact rigorously we need the
following lemma, similar to Lemma 9 in \cite{FGS3}.
\begin{lemma} Assume that Hypothesis (H1) is satisfied and that $R' >R >0$.
Then, for every $\mu \geq 0$, there exists a constant $C_{\mu}$
such that
\begin{equation}\label{eq:decay}
\sup_{X,x \in \R^3} e^{-\alpha |x|} \chi (|X| \leq R) \,  \| \chi
(|y| \geq R') G_{X,x} \| \leq C_{\mu}  (R'-R)^{-\mu}\,.
\end{equation}
Moreover, if $\Sigma < \Sigma_{\text{ion}}$, we have
\begin{equation}\label{eq:dec2}
\| \phi (\chi (|y| \geq R') G_{X,x}) \chi (|X| \leq R) E_{\Sigma}
(\Hmod) \| \leq C_{\mu} (R'-R)^{-\mu}
\end{equation}
\end{lemma}
{\it Remark.} In the proof of the existence of the operators $W$
and $W_+$, where we use this lemma, typically $R= \beta t$ and
$R'= \gamma t$. Hence the r.h.s. of (\ref{eq:dec2}) gives a decay
in time which is integrable if we choose $\mu$ large enough.
\begin{proof}
To prove (\ref{eq:decay}), we first choose $\eps = (R'-R)/2\lambda
>0$, with $\lambda = \max (\lambda_n,\lambda_e)$ (recall that
$\lambda_e = m_n/M$ and $\lambda_n = m_e/M$) and we observe that
\begin{equation}\label{eq:decay1}
\begin{split}
e^{-\alpha|x|} \chi (|X| \leq R) \| \chi (|y| \geq R') G_{X,x} \| 
\leq \; &\chi (|x| \leq \eps) \, \chi (|X| \leq R)
\, \| \chi (|y| \geq R') G_{X,x} \|
\\ &+ e^{-\alpha \eps } \,  \| G_{X,x} \| \,.
\end{split}
\end{equation}
Using that $G_{X,x} (k) = e^{-i(X+\lambda_e x) \cdot k} \kappa_e
(k) + e^{-i(X-\lambda_n x) \cdot k} \kappa_n (k)$, it follows that
\begin{equation}
\| G_{X,x} \|^2 \leq 2 \int \rd k \, \left( |\kappa_e (k)|^2 +
|\kappa_n (k)|^2 \right).
\end{equation}
Hence the second term on the r.h.s. of (\ref{eq:decay1}) can be
bounded by $C \eps^{-\mu}= C(R'-R)^{-\mu}$ (because $\eps^{\mu}
e^{-\alpha \eps}$ is bounded). Moreover the square of the first
term on the r.h.s. of (\ref{eq:decay1}) can be estimated by
\begin{equation}\label{eq:declast}
\begin{split}
\chi (|x| \leq &\eps) \, \chi (|X| \leq R) \, \| \chi (|y| \geq
R') G_{X,x} \|^2 \\ \leq \; &2 \, \chi (|x| \leq \eps) \, \chi
(|X| \leq R) \int \rd y \,\chi (|y| \geq R') \, \left(
|\hat{\kappa}_e (X+\lambda_e x - y)|^2 + |\hat{\kappa}_n
(X-\lambda_n x -y)|^2 \right) \\ \leq \; &2 \int \rd y \, \chi
\left(|X+\lambda_e x -y| \geq \frac{R'-R}{2} \right) \, |\hat{\kappa}_e
(X+\lambda_e x -y)|^2
\\ & + 2 \int \rd y \, \chi \left( |X-\lambda_n x -y| \geq
\frac{R'-R}{2} \right) \, |\hat{\kappa}_n (X-\lambda_n x -y)|^2 \\
\leq  &\;C \int_{|y| \geq \frac{R'-R}{2}} \rd y \left(
|\hat{\kappa}_n (y)|^2 + |\hat{\kappa}_e (y)|^2 \right)
\end{split}
\end{equation}
for all $X$ and $x$. Here we used that, from $|y| \geq R'$, $|X|
\leq R$, and since, by definition of $\eps$, $\lambda_e |x| \leq
\lambda_e \eps \leq (R'-R)/2$ and $\lambda_n |x| \leq \lambda_n
\eps \leq (R'-R)/2$, we have that $|X+\lambda_e x -y| \geq |y| -
|X| - \lambda_e |x| \geq (R'-R)/2$ and analogously $|X-\lambda_n
-y| \geq (R'-R)/2$. Since $\kappa_e,\kappa_n \in C_0^{\infty}
(\R^3)$, their Fourier transform decay faster than any power, and
hence (\ref{eq:declast}) implies (\ref{eq:decay}).

To prove (\ref{eq:dec2}), we use that
\begin{equation}
\begin{split}
\| \phi (\chi &(|y| \geq R') G_{X,x}) \chi (|X| \leq R) E_{\Sigma}
(\Hmod) \| \\&\leq \| e^{-\alpha |x|} \chi (|X| \leq R) \phi (\chi
(|y| \geq R') G_{X,x}) (N+1)^{-1} \| \| (N+1) e^{\alpha |x|}
E_{\Sigma} (\Hmod)\| \\
&\leq C \sup_{x,X} e^{-\alpha|x|} \chi (|X| \leq R) \| \chi (|y|
\geq R') G_{X,x} \|
\end{split}
\end{equation}
because $\| (N+1) e^{\alpha |x|} E_{\Sigma} (\Hmod)\|$ is finite
(because $\Sigma < \Sigma_{\text{ion}}$ and by a simple
commutation). Eq. (\ref{eq:dec2}) then follows from
(\ref{eq:decay}).
\end{proof}

The decay of the interaction determined in the last lemma is one
of the two key ingredients for proving the existence of $W$ and
$W_+$. The other one is a propagation estimate for the photons,
analogous to Proposition 24 in \cite{FGS3}, but with the cutoff
for $x/t$ (in \cite{FGS3}, $x$ is the position of the electron)
replaced by a cutoff for the asymptotic velocity $X/t$ of the
center of mass of the nucleus-electron compound (the reason why we
can introduce here a cutoff in $X/t$ is that, because of
(\ref{eq:PEX}), we know it can not exceed $\beta$).

For the details of the proof of the next two proposition we refer
to Theorems 26 and 28 in \cite{FGS3}.

\begin{prop}[Existence of the asymptotic observable]\label{prop:W}
Assume that Hypotheses (H0), (H1) and (H3) are satisfied. Fix
$\beta < 1$, and choose $\Sigma < \min (\Sigma_{\beta} ,
\Sigma_{\text{ion}})$. Suppose that \(f\in C_0^{\infty}(\R)\) with
\(\supp(f)\subset(-\infty,\Sigma)\). Let $\gamma$, and
$\chi_{\gamma}$ be as defined above, and let $\chi_{\gamma,t}$ be
the operator of multiplication with $\chi_{\gamma}(|y|/t)$. Then,
for $g \geq 0$ small enough (in order for (\ref{eq:Sigma1}) to
hold true),
\begin{equation*}
W = s-\lim_{t\to \infty} e^{i\Hmod t}f\dGamma(\chi_{\gamma,t})f
e^{-i\Hmod t}
\end{equation*}
exists, $W=W^{*}$ and $W$ commutes with $\Hmod$. Here
$f=f(\Hmod)$.
\end{prop}

\begin{prop}[Existence of $W_{+}$]\label{prop:W+}
Assume Hypotheses (H0), (H1) and (H3) are satisfied. Fix $\beta <
1$ and choose $\Sigma < \min (\Sigma_{\beta},
\Sigma_{\text{ion}})$. Suppose that \(f\in C_0^{\infty}(\R)\) with
\(\supp(f)\subset (-\infty,\Sigma)\), and that $\chi_{\gamma}$ and
$j_t$ are defined as described above. If $g \geq 0$ is so small
that (\ref{eq:Sigma1}) holds, then
\begin{itemize}
\item[(i)] the limit
\[  W_{+} = s-\lim_{t\to\infty} e^{i\Hmodex t}  f (\Hmodex)
\uGamma(j_t)\dGamma(\chi_{\gamma,t}) f (\Hmod) e^{-i\Hmod t} \]
exists, and \(e^{-i\Hmodex s}W_{+} = W_{+}e^{-i\Hmod s}\), for all
$s\in \R$; \item[(ii)] \( ( 1\otimes \chi(N=0)) W_{+} = 0\);
\item[(iii)] \(W=\tilde \Omega_{+} W_{+}\).
\end{itemize}
\end{prop}

\subsection{Positivity of the Asymptotic Observable and Asymptotic Completeness}
\label{sec:pos}

In this section we prove the positivity of the asymptotic
observable $W$, restricted to the subspace of states orthogonal to
wave packets of dressed atom states and not containing any soft
bosons. We need the following lemma.

\begin{lemma}\label{lm:poscomm3}
Assume Hypotheses (H0), (H1), (H3). Fix $\beta < 1$ and choose
$\Sigma < \min (\Sigma_{\beta}, \Sigma_{\text{ion}})$. Suppose,
moreover, that $f\in C_0^{\infty}(\R)$ and
\(\supp(f)\subset(-\infty,\Sigma)\). Put $a_X =(1/2) (\nabla
\omega \cdot (y-X) + (y - X) \cdot \nabla \omega)$, where $X$ is
the position of the center of mass of the atom. Then, if $g$ is so
small that (\ref{eq:Sigma1}) holds true, we have that
\begin{equation}
f(\Hmod)[i\Hmod,\dGamma(a_X)]f(\Hmod) \geq (1-\beta) f(\Hmod) N
f(\Hmod) - C g f(\Hmod)^2
\end{equation}
on the range of the projection $\Gamma (\chi_i)$.
\end{lemma}

This lemma follows from a straightforward estimate of the
commutator $[\Hmod, \dGamma (a_X)]$, from (\ref{eq:Sigma1}),
(\ref{eq:Sigma2}), and from Lemma \ref{lm:phibounds}.

\begin{theorem}[Positivity of the asymptotic observable]
\label{thm:Wpos} Assume that Hypotheses (H0)-(H3) are satisfied.
Fix $\beta <1/3$ and choose $\Sigma < \min
(\Sigma_{\beta},\Sigma_{\text{ion}})$ (with $\Sigma_{\beta}$ and
$\Sigma_{\text{ion}}$ as in Lemma \ref{lm:Sigma}). 
Let the operator $W$ be defined as in
Proposition \ref{prop:W} with $\supp f \subset (-\infty,\Sigma)$.
Then if $g
>0$ is sufficiently small we can choose $\gamma > \beta$ in the
definition of $W$ such that
\begin{equation}\label{eq:claim}
\sprod{\ph}{W\ph} \geq C \| f (\Hmod) \ph \|^2, \quad \quad
\text{for all } \ph \in \ran \Pgs^{\perp} \Gamma (\chi_i) \, .
\end{equation}
Here $C$ is a positive constant depending on $g$, $\beta$,
$\Sigma$, but independent of $\ph$. In particular, if $\Delta
\subset (-\infty, \Sigma)$ and then $f=1$ on $\Delta$, then
\begin{equation}
W|_{\ran E_{\Delta}(\Hmod)\Gamma (\chi_i) P_{\text{das}}^{\perp}}
\geq C > 0.
\end{equation}
\end{theorem}
\begin{proof}
Let $\mathcal{D} = D( \dGamma (a)) \cap \ran \Gamma (\chi_i)
\Pgs^{\perp}$. Since $\mathcal{D}$ is dense in $\ran \Gamma
(\chi_i)\Pgs^{\perp}$, it is enough to prove that
\begin{equation}\label{eq:phWph}
\sprod{\ph}{W\ph} \geq C \| f \ph \|^2
\end{equation}
for every $\ph \in \mathcal{D}$. As before, we use the notation $f
= f(\Hmod)$.

The first step consists in proving that there exists a constant
$C$, depending only on $\Sigma$, such that, for every $\ph \in
\mathcal{D}$ and for every $\eps >0$,
\begin{equation}\label{eq:step1+2}
\begin{split}
\sprod{\ph_{t}}{f \dGamma (\chi_{\gamma,t}) f \ph_{t}} \geq
 \; &C \| f\ph \|^{-2} \left[ \frac{1-\beta}{t} \int_0^t \rd s
 \sprod{\ph_{s}}{f N f \ph_{s}} - (\gamma + \beta + \eps) \sprod{\ph_t}{f Nf \ph_t}
 \right]^2 \\ &- C g \|f\ph\|^2 + o(1),\qquad \text{as }\quad t
 \to\infty\, .
\end{split}
\end{equation}
This inequality can be established as in the proof of Theorem 27
in \cite{FGS3}. Next, we observe that
\begin{equation}\label{eq:fNf}
\frac{1}{t} \int_0^t \rd s \sprod{\ph_s}{fNf\ph_s} \geq \| f\ph
\|^2 - \frac{1}{t} \int_0^t \sprod{\ph_s}{f P_{\Omega} f \ph_s}
\end{equation}
where $P_{\Omega}$ denotes the orthogonal projection onto the Fock
vacuum $\Omega$. The second term on the r.h.s. of the last
equation can be written as an integral over fibers with fixed
total momentum. Making use of the fact that $\ph = \Pgs^{\perp}
\ph$ and of Fubini's Theorem to interchange the integration over
$s$ and over $\Pi$, we obtain
\begin{equation}\label{eq:int}
\frac{1}{t} \int_0^t \rd s \, \sprod{\ph_s}{f P_{\Omega} f \ph_s}
= \int \rd \Pi \, \frac{1}{t} \int_0^t \rd s \, \| P_{\Omega} f
(\Hmod (\Pi))e^{-i\Hmod (\Pi) s} P_{\psi_{\Pi}}^{\perp} \ph (\Pi)
\|^2 \,
\end{equation}
where $P_{\psi_{\Pi}}= |\psi_{\Pi} \rangle \langle \psi_{\Pi}|$ is
the orthogonal projection onto the dressed atom state
$\psi_{\Pi}$, and $P_{\psi_{\Pi}}^{\perp} = 1 - P_{\psi_{\Pi}}$ is
its orthogonal complement. For every fixed $\Pi$, the operator
$P_{\Omega} f (\Hmod (\Pi))$ is a compact operator on $L^2
(\R^3,\rd x) \otimes \F$, because $\| e^{\alpha |x|} E_{\Sigma}
(\Hmod (\Pi)) \| \leq C$; (since $\Sigma < \Sigma_{\text{ion}}$,
this follows from Lemma \ref{lm:Sigma}). By the continuity of the
spectrum of $\Hmod (\Pi)$ on $\ran E_{\Sigma} (\Hmod)
P_{\psi_{\Pi}}^{\perp} \Gamma (\chi_i)$ (see Corollary
\ref{cor:spectrum}), and the RAGE Theorem (see, for example,
\cite{RS3}), it follows that
\begin{equation}
\frac{1}{t} \int_0^t \rd s \,  \| P_{\Omega} f (\Hmod
(\Pi))e^{-i\Hmod (\Pi) s} P_{\psi_{\Pi}}^{\perp} \ph (\Pi) \|^2
\to 0,
\end{equation}
as $t \to \infty$, pointwise in $\Pi$. Using Lebesgue's Dominated
Convergence Theorem, we conclude that
\begin{equation}\label{eq:fPf}
\frac{1}{t} \int_0^t \rd s \, \sprod{\ph_s}{f P_{\Omega}f \ph_s}
\to 0 \, ,
\end{equation}
as $t \to \infty$. {F}rom (\ref{eq:fNf}) we obtain
\begin{equation}\label{eq:fNf2}
\frac{1}{t} \int_0^t \rd s \, \sprod{\ph_s}{f N f \ph_s} \geq
\frac{\| f \ph \|^2}{2}
\end{equation}
for $t$ large enough, where we can assume $\| f \ph \| \neq 0$
without loss of generality. Eqs. (\ref{eq:fPf}) and
(\ref{eq:fNf2}) allow us to apply Lemma \ref{lm:teilfolge}, with
$h_1 (s) = \sprod{\ph_s}{fNf\ph_s}$ and $h_2 (s) =
\sprod{\ph_s}{fP_{\Omega} f \ph_s}$ (it is easy to check that
$h_1$ and $h_2$ are bounded and continuous). We conclude that
there exists a sequence $\{ t_n \}_{n \geq 0}$ with $t_n \to
\infty$, as $n \to \infty$, such that
\begin{equation}
\begin{split}\label{eq:teilfolge}
\frac{1}{t_n} \int_0^{t_n} \rd s \, \sprod{\ph_s}{fNf \ph_s} &\geq
(1 -\eps) \sprod{\ph_{t_n}}{ fNf \ph_{t_n}}\, , \quad \quad
\text{and}
\\ \sprod{\ph_{t_n}}{ f P_{\Omega} f \ph_{t_n}} &\to 0 \, , \quad
\text{as } n \to \infty \, .
\end{split}
\end{equation}
{F}rom (\ref{eq:step1+2}) we infer that
\begin{equation}
\sprod{\ph_{t_n}}{f \dGamma (\chi_{\gamma,t_n}) f \ph_{t_n}} \geq C
\| f\ph \|^{-2} ( 1- 2 \beta- \gamma - 2 \eps)^2
\sprod{\ph_{t_n}}{f Nf \ph_{t_n}}^2 - C g \|f\ph\|^2 + o(1)\, ,
\end{equation}
as $n \to \infty$. Choosing $\gamma-\beta$ and $\eps >0$
sufficiently small, we conclude that
\begin{equation}
\begin{split}
\sprod{\ph_{t_n}}{f \dGamma (\chi_{\gamma,t_n}) f \ph_{t_n}} &\geq C
\frac{(1-3\beta)^2}{2} \| f \ph \|^{-2} \sprod{\ph_{t_n}}{f Nf
\ph_{t_n}}^2 - C g \|f\ph\|^2 + o(1)
\\ & \geq C \frac{(1-3\beta)^2}{2} \left( \| f \ph \|^2 - 2
\sprod{\ph_{t_n}}{f P_{\Omega} f \ph_{t_n}} \right) - C g \| f \ph
\|^2 + o(1),
\end{split}
\end{equation}
as $n \to \infty$. Hence, by (\ref{eq:teilfolge}), there are
constants $C_1
>0$ and $C_2<\infty$, depending only on $\Sigma$, such that
\begin{equation}
\sprod{\ph_{t_n}}{f \dGamma (\chi_{\gamma,t_n}) f \ph_{t_n}} \geq
C_1 ( 1-3\beta - C_2 g)^2 \| f \ph \|^2 + o(1) \, ,
\end{equation}
as $n \to \infty$. If $\beta <1/3$ and $g$ is small enough, we
arrive at (\ref{eq:claim}) by taking the limit $n \to \infty$.
(Since we already know that the limit defining $W$ exists, it is
enough to prove its positivity on some arbitrary subsequence!)
\end{proof}

\begin{lemma}\label{lm:teilfolge}
Suppose $h_1$ and $h_2$ are positive, continuous, bounded
functions on $\R$, such that
\begin{equation}\label{eq:ass1}
m_1 (t) := \frac{1}{t} \int_0^t \rd s \, h_1 (s) \geq C >0
\end{equation}
for all $t >0$ large enough, and
\begin{equation}\label{eq:ass2}
m_2 (t) := \frac{1}{t} \int_0^t \rd s h_2 (s) \to 0 \quad \text{as
} \quad t \to \infty \, .
\end{equation}
Then, for every $\delta >0$, there exists a sequence $\{ t_n \}_{n
\geq 0}$, with $t_n \to \infty$, as $n \to \infty$, such that
\begin{equation}\label{eq:cond1}
m_1 (t_n) \geq \frac{1}{1+ \delta}\,  h_1 (t_n)
\end{equation}
and \begin{equation}\label{cond2} h_2 (t_n) \to 0 \quad \quad
\text{as } n \to \infty. \end{equation}
\end{lemma}
\begin{proof}
Define the sets
\[ S_T := \{ t \in [0,T] : m_1 (t) < \frac{1}{1+ \delta} h_1 (t) \}
\, , \quad \quad \text{for some } \delta>0.
\] By the continuity of $h_1 (t)$ and $m_1 (t)$ the sets $S_T$ are
measurable (with respect to Lebesgue measure on $\R$), for all
$T$. Denote by $\mu (A)$ the Lebesgue measure of a measurable set
$A \subset \R$. We show that
\begin{equation}\label{eq:wrong}
\liminf_{T\to\infty} \frac{\mu (S_T)}{T} < 1 .
\end{equation}
In fact, if (\ref{eq:wrong}) were false, then (since $\mu (S_T)/T
\leq 1$ for all $T\geq 0$) \[ \lim_{T \to \infty} \frac{\mu
(S_T)}{T} =1 \] and hence, for arbitrary $\eps
>0$, we could find a $T_0$ such that
\[ \frac{\mu (S_T)}{T} \geq 1- \eps \, , \] for all $T >T_0$. This
would imply that
\begin{equation}
\begin{split}
m_1 (T) &= \frac{1}{T} \int_0^T \rd s \, h_1 (s) \geq \frac{1}{T}
\int_{S_T} \rd s \, h_1 (s) \\ &\geq \frac{1+\delta}{T} \int_{S_T}
\rd s \, m_1 (s) \geq \frac{1+\delta}{T} \int_0^T \rd s \, m_1 (s)
- \frac{\mu (S_T^c)}{T} \| m_1 \|_{\infty}
\end{split}
\end{equation}
where $\| m_1 \|_{\infty}$ denotes the supremum of the bounded
function $m_1$ and $S_T^c = [0,T] \backslash S_T$ is the
complement of $S_T$ inside $[0,T]$. Hence, we find
\begin{equation}
m_1 (T) \geq \frac{1+\delta}{T} \int_0^T \rd s \, m_1 (s) - \eps
\|m_1\|_{\infty}
\end{equation}
for every $T \geq T_0$. Put $\widetilde{m}_1 (T) := (1/T) \int_0^T
\rd s \, m_1 (s)$. Then we have
\begin{equation}
\frac{d}{dT} \log \widetilde{m}_1 (T) =
\frac{\widetilde{m}'_1(T)}{\widetilde{m}_1 (T)} = \frac{1}{T}
\left( \frac{m_1 (T)}{\widetilde{m}_1 (T)} -1 \right) \geq
\frac{1}{T} \left(\delta - \frac{\eps
\|m_1\|_{\infty}}{\widetilde{m}_1 (T)}\right) \, .
\end{equation}
By the assumption that $m_1 (T) \geq C$ for all $T$ large enough,
we have $\widetilde{m}_1 (T) \geq C$, and thus, choosing $\eps < C
\delta /2 \|m_1\|_{\infty}$, we find
\begin{equation}
\frac{d}{dT} \log \widetilde{m}_1 (T) \geq \frac{\delta}{2 T}
\end{equation}
for all $T$ large enough. This contradicts the boundedness of
$\widetilde{m}_1 (T)$ (which follows from the boundedness of
$m_1(T)$). This proves (\ref{eq:wrong}), and implies that there
exist $\eps >0$ and a sequence $\{ T_m \}_{m \geq 0}$ converging
to infinity such that
\begin{equation}\label{eq:mass}
\frac{\mu (S_{T_m})}{T_m} \leq 1 - \eps \, ,
\end{equation}
for all $m \geq 0$. Hence $\mu (S_{T_m}^c) \geq \eps T_m$, for all
$m$. Next, we show that there exists a sequence $\{t_n\}_{n \geq
0}$, with $t_n \to \infty$ as $n \to \infty$, such that $t_n \in
\cup_{m \geq 0} S_{T_m}^c$, for all $n \geq 0$, and
\begin{equation}\label{eq:cond2}
h_2 (t_n) \to 0\, ,
\end{equation}
as $n \to \infty$. Since, for all $n \geq 0$, $t_n \in S_{T_m}^c$,
for some $m \in \N$, the sequence $t_n$ automatically satisfies
(\ref{eq:cond1}). Thus the lemma follows if we can prove
(\ref{eq:cond2}). To this end we argue again by contradiction. If
there were no sequence $\{ t_n \}_{n \geq 0} \in \cup_{m \geq 0}
S_{T_m}^c$ satisfying (\ref{eq:cond2}) then there would exist
$\tau$ and $\alpha >0$ such that $h_2 (t) \geq \alpha$, for all $t
\in \cup_{m \geq 0} S_{T_m}^c \cap [\tau, \infty)$. But then, for
an arbitrary $m \in \N$ with $T_m \geq \tau$,
\begin{equation}
\begin{split}
\frac{1}{T_m} \int_0^{T_m} \rd s \, h_2 (s) &\geq \frac{1}{T_m}
\int_{S_{T_m}^c \cap [\tau, T_m]} \rd s \, h_2 (s) \geq \alpha
\frac{\mu (S_{T_m}^c \cap [\tau, T_m])}{T_m} \\ &\geq \alpha
\frac{\mu (S_{T_m}^c)}{T_m} - \frac{\alpha \tau}{T_m} \geq \alpha
\eps - \frac{\alpha \tau}{T_m}
\end{split}
\end{equation}
for all $m \in \N$ with $T_m \geq \tau$. Taking $m \to \infty$,
this contradicts the assumption (\ref{eq:ass2}).
\end{proof}

\subsection{Asymptotic Completeness}\label{sec:AC}

Using the positivity of the asymptotic observable $W$, we can
complete the proof of asymptotic completeness for the Hamiltonian
$\Hmod$. Our proof is based on induction in the energy. The
following simple lemma is useful.

\begin{lemma}\label{lm:induction}
Assume that Hypotheses (H0)-(H3) are satisfied. Fix $\beta <1$ and
choose $\Sigma < \min (\Sigma_{\beta}, \Sigma_{\text{ion}})$. The
wave operators $\tilde{\Omega}_+$ and $\Omega_+$ are defined as in
Lemma \ref{lm:tilde-Omega-mod} and in Theorem~\ref{thm:Omega_+},
respectively. Suppose that $\ran \Omega_+ \supset E_{\eta} (\Hmod)
\H$, for some $\eta < \Sigma$. Then, for every $\ph \in \ran
E_{\Sigma} (\Hmodex)$, there exists $\psi \in \ran E_{\Sigma}
(\Hmodex)$ such that
\[\tilde{\Omega}_{+}(E_{\eta}(\Hmod)\otimes 1)\ph = \Omega_{+}\psi.\]
If $\Delta \subset ( -\infty , \Sigma)$ and \(\ph\in
E_{\Delta}(\Hmodex)\HxF\) then \(\psi\in
E_{\Delta}(\Hmodex)\HxF\).
\end{lemma}

The interpretation of this lemma is simple: If we know that
asymptotic completeness holds for vectors with energy lower than
$\eta$, then it continues to be true if we add asymptotically free
photons to these vectors (no matter what the total energy of the
new state is). For the proof of this lemma we refer to Lemma 20 of
\cite{FGS3}. Using this lemma we can prove asymptotic completeness
for $\Hmod$; the proof is similar to the proof of Theorem
\ref{thm:AC} in \cite{FGS3}. We repeat it here, because it is very
short, and because it explains the ideas behind all the tools
introduced in Section \ref{sec:scatt}.

\begin{theorem}\label{thm:AC}
Assume that Hypotheses (H0)-(H3) are satisfied. Fix $\beta < 1/3$
and choose $\Sigma <\min (\Sigma_{\beta}, \Sigma_{\text{ion}})$;
(with $\Sigma_{\beta}$ and $\Sigma_{\text{ion}}$ defined as in
Lemma \ref{lm:Sigma}). If $g
> 0$ is sufficiently small, then
\[ \Ran \Omega_+ \supset E_{(-\infty , \Sigma)} (\Hmod) \H . \]
\end{theorem}

\begin{proof}
The proof is by induction in energy steps of size $m=\sigma/2$.
We show that
\begin{equation}\label{eq:to-prove}
\Ran \Omega_{+} \supset E_{(-\infty,\Sigma-km)}(\Hmod) \H 
\end{equation}
holds for $k=0$, by proving this claim for all \(k\in
\{0,1,2,\ldots\}\). Since $\Hmod$ is bounded below,
\eqref{eq:to-prove} is obviously correct for $k$ large enough.
Assuming that \eqref{eq:to-prove} holds for $k=n+1$, we now prove
it for $k=n$. Since $\Ran \Omega_{+}$ is closed (by
Theorem~\ref{thm:Omega_+}) and since $\ran \Omega_+ \supset \Hgs$,
it suffices to prove that
\begin{equation*}
  \Ran{\Omega_{+}} \supset \Pgs^{\perp} \Gamma (\chi_i)
  E_{\Delta}(\Hmod)\H \, ,
\end{equation*}
for $\Delta = (\inf \sigma (H_{g=0}) -1 ,\Sigma -nm)$. Here we use
Lemma~\ref{lm:AC}. Fix $\widetilde\Sigma$ with $\Sigma <
\widetilde\Sigma < \min (\Sigma_{\beta},\Sigma_{\text{ion}})$ and
choose \(f\in C_0^{\infty}(\R)\) real-valued, with $f\equiv 1$ on
$\Delta$ and \(\supp(f)\subset(-\infty,\widetilde \Sigma)\). We
define the asymptotic observable $W$ in terms of $f$, as in
Proposition~\ref{prop:W}. By Theorem~\ref{thm:Wpos}, the operator
\(\Gamma(\chi_i)\Pgs^{\perp} W \Pgs^{\perp} \Gamma (\chi_i)\) is
strictly positive on \(\Pgs^{\perp} \Gamma (\chi_i)
E_{\Delta}(\Hmod)\H\), and hence onto, if $g$ is small
enough. Given $\psi$ in this space, we can therefore find a vector
\(\ph = \Pgs^{\perp} \Gamma (\chi_i)\ph\) such that
\begin{equation*}
  \Pgs^{\perp} \Gamma (\chi_i) W\ph =\psi.
\end{equation*}
By Proposition \ref{prop:W+}, \(W\ph=\tilde \Omega_{+}W_{+}\ph\)
and \(W_{+}\ph = E_{\Sigma-nm}(\Hmodex)W_{+}\ph\). Furthermore, by
part (ii) of Proposition~\ref{prop:W+}, $W_{+}\ph$ has at least
one boson in the outer Fock space, and thus an energy of at most
\(\Sigma-(n+1)m\) in the inner one. That is,
\begin{equation*}
  W_{+}\ph = [E_{\Sigma-(n+1)m}(\Hmod)\otimes 1]W_{+}\ph .
\end{equation*}
Hence we can use the induction hypothesis \(\Ran \Omega_{+}
\supset E_{\Sigma-(n+1)m}(\Hmod)\H\). By Lemma~\ref{lm:induction},
it follows that \(\tilde \Omega_+ W_{+}\ph = \Omega_{+}\gamma\)
for some \(\gamma\in E_{\Delta}(\Hmodex)\H\). We conclude that
\begin{eqnarray*}
  \psi &=& \Gamma(\chi_i) \Pgs^{\perp} \Omega_{+} \gamma\\
       &=& \Gamma(\chi_i) \Omega_{+} (1\otimes P_{\Omega}^{\perp})\gamma\\
       &=& \Omega_{+} (\Gamma(\chi_i)\otimes \Gamma(\chi_i) P_{\Omega}^{\perp})\gamma,
\end{eqnarray*}
where $P_{\Omega}^{\perp}$ is the projection onto the orthogonal
complement of the vacuum. This proves the theorem.
\end{proof}

\appendix

\section{Fock Space and Second Quantization}
\label{sec:Fock}

Let $\h$ be a complex Hilbert space, and let $\otimes_s^n\h$
denote the $n$-fold symmetric tensor product of $\h$. Then the
bosonic Fock space over $\h$,
\[  \F=\F(\h)=\bigoplus_{n\geq0}\h^{\otimes_s n} \, ,\]
is the space of sequences \(\ph=(\ph_n)_{n\geq 0}\), with
$\ph_0\in \C$, \(\ph_n\in \otimes_s^n\h\), and with the scalar
product given by
\[ \sprod{\ph}{\psi} := \sum_{n\geq 0} ( \ph_n  , \psi_n ), \]
where \(( \ph_n,\psi_n)\) denotes the inner product in
\(\otimes^n_s \h\). The vector \(\Omega=(1,0,\ldots)\in\F\) is
called the vacuum. By $\F_0\subset\F$ we denote the dense subspace
of vectors $\ph$ for which $\ph_n=0$, for all but finitely many
$n$. The number operator $N$ is defined by \((N\ph)_n=n\ph_n\).

\subsection{Creation- and Annihilation Operators}

The creation operator $a^*(h)$, $h\in\h$, is defined on
$\h^{\otimes_s n-1}$ by
\[ a^*(h)\ph = \sqrt{n}\,S(h\otimes \ph),\hspace{3em}\mbox{for}\
\ph\in \h^{\otimes_s n-1} ,\] and extended by linearity to $\F_0$.
Here $S$ denotes the orthogonal projection onto the symmetric
subspace \(\otimes_s^n\h\subset \otimes^n\h\). The annihilation
operator $a(h)$ is the adjoint of $a^*(h)$. Creation- and
annihilation operators satisfy the canonical commutation relations
(CCR)
\begin{equation*}
[a(g),a^{*}(h)] = (g,h),\hspace{3em} [a^{\#}(g),a^{\#}(h)] =0.
\end{equation*}
In particular, \([a(h),a^{*}(h)] = \|h\|^2\), which implies that
the graph norms associated with the closable operators $a(h)$ and
$a^{*}(h)$ are equivalent. It follows that the closures of $a(h)$
and $a^{*}(h)$ have the same domain. On this common domain we
define the self-adjoint operator
\begin{equation}\label{eq:phi}
\phi(h) = a(h) + a^{*}(h).
\end{equation}
The creation- and annihilation operators, and thus $\phi(h)$, are
bounded relative to the square root of the number operator:
\begin{equation}\label{eq:aN_bound}
\|a^{\#}(h)(N+1)^{-1/2}\| \leq \|h\| \, .
\end{equation}
More generally, for any $p\in\R$ and any integer $n$,
\begin{equation*}
\|(N+1)^pa^{\#}(h_1)\ldots a^{\#}(h_n)(N+1)^{-p-n/2}\| \leq
C_{n,p}\,\|h_1\|\cdot\ldots\cdot\|h_n\|.
\end{equation*}

\subsection{The Functor $\Gamma$}\label{sec:A2}

Let $\h_1$ and $\h_2$ be two Hilbert spaces and let
\(b\in\L(\h_1,\h_2)\). We define $\Gamma(b)\ :\
\F(\h_1)\rightarrow\F(\h_2)$ by
\begin{equation*}
\Gamma(b)\restricted\otimes_s^n \h_{1} = b\otimes\ldots\otimes b.
\end{equation*}
In general $\Gamma(b)$ is unbounded; but if $\|b\|\leq 1$ then
\(\|\Gamma(b)\|\leq 1\). From the definition of $a^{*}(h)$ it
easily follows that
\begin{alignat}{2}
\Gamma(b)a^{*}(h) & = a^{*}(bh)\Gamma(b),\qquad  & h\in\h_1&
\label{geq1}\\ \Gamma(b)a(b^*h) &= a(h)\Gamma(b),   &
h\in\h_2&.\label{geq2}
\end{alignat}
If $b^{*}b=1$ on $\h_1$ then these equations imply that
\begin{alignat}{2}
\Gamma(b)a(h) & = a(bh)\Gamma(b)\qquad & h\in\h_1 \label{geq3}&\\
\Gamma(b)\phi(h) & = \phi(bh)\Gamma(b) & h\in\h_1 \label{geq4}&.
\end{alignat}

\subsection{The Operator $\dGamma(b)$}

Let $b$ be an operator on $\h$. Then $\dGamma(b)\ :\
\F(\h)\rightarrow\F(\h)$ is defined by
\begin{equation*}
\dGamma(b)\restricted\otimes_s^n\h =\sum_{i=1}^n(1\otimes\ldots
b\otimes\ldots 1).
\end{equation*}
For example $N=\dGamma(1)$. From the definition of $a^{*}(h)$ we
infer that
\begin{equation*}
[\dGamma(b),a^*(h)] =  a^*(bh) \quad [\dGamma(b),a(h)] =
-a(b^*h),
\end{equation*}
and, if $b=b^{*}$,
\begin{equation}\label{eq:dgamma-phi}
i[\dGamma(b),\phi(h)]  = \phi(ibh).
\end{equation}
Note that \(\|\dGamma(b)(N+1)^{-1}\|\leq  \|b\|\).

\subsection{The Tensor Product of two Fock Spaces}
\label{sec:FxF} Let $\h_1$ and $\h_2$ be two Hilbert spaces. We
define a linear operator
\(U:\F(\h_1\oplus\h_2)\rightarrow\F(\h_1)\otimes\F(\h_2)\) by
\begin{equation}\label{ueq0}
\begin{split}
U\Omega &= \Omega\otimes\Omega\\
U a^*(h) &= [a^*(h_{(0)})\otimes 1+ 1\otimes
a^*(h_{(\infty)})]U\hspace{3em}\text{for
}h=(h_{(0)},h_{(\infty)})\in\h_1\oplus\h_2.
\end{split}
\end{equation}
This defines $U$ on finite linear combinations of vectors of the
form \(a^*(h_1)\ldots a^*(h_n)\Omega\). From the CCRs it follows
that $U$ is isometric. Its closure is isometric and onto, hence
unitary.

\subsection{Factorizing Fock Space in a Tensor Product}
\label{sec:factfock}

Suppose $j_0$ and $j_{\infty}$ are linear operators on $\h$ and
\(j:\h\rightarrow\h\oplus\h \) is defined by
\(jh=(j_0h,j_{\infty}h),\ h\in\h\). Then
\(j^*(h_1,h_2)=j_0^*h_1+j_{\infty}^*h_2\) and consequently
\(j^*j=j_0^*j_0+j_{\infty}^*j_{\infty}\).
We define
\[ \uGamma(j)=U\Gamma(j):\F\rightarrow\F\otimes\F \, ,\]
where $\Gamma (j)$ is as defined in Sect. \ref{sec:A2}. It follows
that \(\uGamma(j)^*\uGamma(j)=\Gamma(j^*j)\) which is the identity
if $j^*j=1$. In this case
\begin{align}\label{eq:ugamma-a}
\uGamma(j)a^{\#}(h) & = [a^{\#}(j_0h)\otimes 1+ 1\otimes
a^{\#}(j_{\infty}h)]\uGamma(j)\\
\label{eq:ugamma-phi}\uGamma(j)\phi(h) & = [\phi(j_0 h)\otimes 1+
1\otimes \phi(j_{\infty}h)]\uGamma(j).
\end{align}

\subsection{The "Scattering Identification"}
\label{sec:scattiden}

We define the scattering identification \( I:\F\otimes\F\to \F\)
by
\begin{align*}
I (\ph\otimes\Omega) &= \ph\\ I \ph\otimes a^*(h_1)\cdots
a^*(h_n)\Omega &= a^*(h_1)\cdots a^*(h_n)\ph,
\hspace{3em}\ph\in\F_0,
\end{align*}
and extend it by linearity to $\F_0\otimes\F_0$. (Note that this
definition is symmetric with respect to the two factors in the
tensor product.) There is a second characterization of $I$ which
can be useful. Let $\iota:\h \oplus \h \to \h$ be defined by
$\iota(h_{(0)},h_{(\infty)})=h_{(0)}+h_{(\infty)}$. Then \(
I=\Gamma(\iota)U^*\), with $U$ as above. Since
$\|\iota\|=\sqrt{2}$, the operator $I$ is unbounded, but it can be
proved that $I (N+1)^{-k} \otimes \chi (N \leq k)$ is bounded, for
any $k\geq 1$.

\section{Bounds on the Interaction} \label{sec:bounds}

In this section we review standard estimates that are used throughout this paper
to bound the interaction.

\begin{lemma}\label{lm:estim} Let \(L^2_{\omega}(\R^3) :=
L^2 (\R^3 , (1 + 1/|k|) dk)\) and let $h\in L^2_{\omega} (\R^3)$.
Then
\begin{eqnarray*}
\| a(h) \ph \| &\leq &  \left( \int dk |h(k)|^2 /|k| \right)^{1/2} \, \| \dGamma (|k|)^{1/2} \ph \| \\
\| a^* (h) \ph \| &\leq & \| h \|_{\omega} \,\| (\dGamma (|k|)+1)^{1/2} \ph \| \\
\| \phi (h) \ph \| &\leq & \sqrt{2} \, \|h \|_{\omega}\,
\| (\dGamma (|k|)+1)^{1/2} \ph \|\\
\pm \phi(h) &\leq & \alpha \dGamma(|k|) +\frac{1}{\alpha}\int dk
\frac{|h(k)|^2}{|k|},\qquad \alpha>0 ,
\end{eqnarray*}
where \( \| h\|_{\omega}^2 = \int dk \, (1 + 1/|k|) |h(k)|^2 \).
\end{lemma}

The next lemma is used to control the factor $\phi (iaF_x)$
appearing in the commutators of Section \ref{sec:poscomm}.

\begin{lemma}\label{lm:phibounds}
Assume Hypothesis (H0)-(H1). Let $a = (1/2) (\hat{k} \cdot y + y
\hat{k})$ with $\hat{k} = k/|k|$ and choose $\Sigma
<\Sigma_{\text{ion}}$. Then there exists $C_{\Sigma} < \infty$
such that
\begin{equation}\label{eq:phi1}
\| \phi (ia F_x) E_{\Sigma} (H_g (\Pi)) \| \leq C_{\Sigma} \, ,
\end{equation}
with $F_x$ as in Eq. (\ref{eq:Fx}). For $a_{X} := (1/2) (\hat{k}
\cdot (y-X) + (y-X) \cdot \hat{k})$,
\begin{equation}\label{eq:phi2}
\| \phi (i a_{X} G_{X,x} ) E_{\Sigma} (H_g) \| \leq C_{\Sigma} \,
,
\end{equation}
where $G_{X,x} (k) = e^{-ik\cdot X} F_x (k)$; (see Eqs.
(\ref{eq:GXx}), (\ref{eq:Fx})).
\end{lemma}
\begin{proof}
Note that
\begin{equation}
\begin{split}
(a F_x)(k) = \; &(i \hat{k} \cdot \nabla_k + 2/|k|) ( e^{- i
\lambda_e k \cdot x} \kappa_e (k) + e^{i\lambda_n k \cdot x} \kappa_n (k) ) \\
= \; &e^{-i\lambda_e k \cdot x} \left( \lambda_e x \cdot \hat{k}
\, \kappa_e (k) + i\hat{k} \cdot \nabla \kappa_e (k) + 2/|k| \kappa_e (k) \right) \\
&+ e^{i\lambda_n k \cdot x} \left( - \lambda_n x \cdot \hat{k}\,
\kappa_n (k) + i\hat{k}\cdot \nabla \kappa_n (k) +2/|k| \kappa_n
(k) \right).
\end{split}
\end{equation}
Eq.~(\ref{eq:phi1}) follows from Lemma \ref{lm:estim}, because
$e^{\alpha|x|} E_{\Sigma} (H_g(\Pi))$ is bounded (see Lemma
\ref{lm:Sigma}) and from Hypothesis (H1). Eq.
(\ref{eq:phi2}) follows from (\ref{eq:phi1}) because
\[ (a_X G_{X, x}) (k) = e^{-i X\cdot k} (a F_x)(k) \, .\]
\end{proof}

\bibliographystyle{alpha}

\end{document}